\documentclass[twocolumn]{aastex631}

\usepackage{rotating}
\usepackage{tablefootnote}
\usepackage{tabularx}
\let\tablewidth\relax
\usepackage{tabularray}

\begin{document}

\title{A systematic analysis of the radio properties of 22 X-ray selected tidal disruption event candidates with the Australia Telescope Compact Array}

\author[0000-0003-3441-8299]{A. J. Goodwin}
\affiliation{International Centre for Radio Astronomy Research – Curtin University, GPO Box U1987, Perth, WA 6845, Australia}
\author{M. Burn}
\affiliation{International Centre for Radio Astronomy Research – Curtin University, GPO Box U1987, Perth, WA 6845, Australia}
\author{G. E. Anderson}
\affiliation{International Centre for Radio Astronomy Research – Curtin University, GPO Box U1987, Perth, WA 6845, Australia}
\author{J. C. A. Miller-Jones}
\affiliation{International Centre for Radio Astronomy Research – Curtin University, GPO Box U1987, Perth, WA 6845, Australia}
\author{I. Grotova}

\affiliation{Max-Planck-Institut für extraterrestrische Physik, Giessenbachstrasse 1, 85748 Garching, Germany}
\author{P. Baldini}
\affiliation{Max-Planck-Institut für extraterrestrische Physik, Giessenbachstrasse 1, 85748 Garching, Germany}
\author{Z. Liu}
\affiliation{Max-Planck-Institut für extraterrestrische Physik, Giessenbachstrasse 1, 85748 Garching, Germany}
\author{A. Malyali}
\affiliation{Max-Planck-Institut für extraterrestrische Physik, Giessenbachstrasse 1, 85748 Garching, Germany}
\author{A. Rau}
\affiliation{Max-Planck-Institut für extraterrestrische Physik, Giessenbachstrasse 1, 85748 Garching, Germany}
\author{M. Salvato}
\affiliation{Max-Planck-Institut für extraterrestrische Physik, Giessenbachstrasse 1, 85748 Garching, Germany}

\begin{abstract}
We present a systematic analysis of the radio properties of an X-ray selected sample of tidal disruption event (TDE) candidates discovered by the eROSITA telescope. 
We find radio sources coincident with half of the transient events (11 TDEs), with 8 radio sources showing statistically significant variability over a 6-month period. We model the radio spectra of 6 sources with sufficiently bright radio emission and find the sources show radio spectra consistent with optically thin synchrotron emission and radio outflow minimum radii of $10^{16}$--$10^{17}$\,cm, velocities 0.01--0.05\,$c$, and energies $10^{48}$--$10^{51}$\,erg. On comparison with the radio properties of an optically-selected TDE sample at similar late times, we find no significant difference in the radio luminosity range or radio detection rate. 
We find a tentative positive trend with peak radio and X-ray luminosity, but require further observations to determine if this is real or due to observational bias due to the large range in distances of the events. 
Interestingly, none of the X-ray selected events show late rising radio emission, compared to 45$\%$ of radio-detected sources of an optically-selected sample that showed late rising radio emission. We propose that this may indicate that many TDEs launch radio outflows at or near peak X-ray luminosity, which can be significantly delayed from peak optical luminosity. This study presents the first systematic analysis of the radio properties of an X-ray selected sample of TDEs, and gives insight into the possible link between the physical processes that power X-ray and radio emission in TDEs.

\end{abstract}


\section{Introduction} \label{sec:intro}
A tidal disruption event (TDE) occurs when a star passes too close to a supermassive black hole (SMBH) and is torn apart by strong tidal forces \citep{Rees1988}. Characterised by X-ray, optical, infra-red, and/or radio flares observed from the nuclei of previously quiescent galaxies, TDEs provide a rare opportunity to observe emission from previously non-active SMBHs in relatively low mass galaxies \citep[e.g.][]{vanVelzen2020}.

The first TDE candidates were discovered at X-ray wavelengths by the ROSAT All Sky Survey \citep{Truemper1982} as soft X-ray outbursts of luminosities $10^{41}-10^{44}$\,erg/s in galaxies that had no previous AGN activity signatures \citep[e.g.][]{Komossa1999,Donley2002}. 
With the advent of sensitive all sky optical surveys, the majority of TDEs now known to date have been discovered by optical flares in previously quiescent galaxies \citep[e.g.][]{Hammerstein2022,Yao2023}. 

TDEs appear to radiate much of their energy in the optical/UV \citep[e.g.][]{Gezari2009,vanVelzen2020} and show transient broad H and/or He lines in their optical spectra \citep[e.g.][]{Yao2023,Hammerstein2022}. The optical emission is consistent with thermal blackbody emission \citep[e.g.][]{Gezari2009,vanVelzen2020} and is thought to be due to an accretion disc \citep[e.g.,][]{Guillochon2014b,Shiokawa2015,Bonnerot2016,Hayasaki2016,Mummery2020}, debris collisions and shocks \citep[e.g.,][]{Dai2015,Lu2020}, or a debris envelope that reprocesses higher energy emission from closer to the SMBH \citep[e.g.,][]{Metzger2016,Roth2016,Metzger2022}. Soft thermal X-ray flares are observed from some events, but can show highly variable behaviour with the X-ray rise sometimes significantly delayed from the initial optical flare \citep[e.g.][]{Kajava2020}, and non-monotonic decays \citep[e.g.][]{Malyali2023b}.

Radio flares are detected in 30--50$\%$ of TDEs, attributed to synchrotron emission produced by material ejected by the TDE interacting with the circumnuclear medium (CNM) \citep[e.g.][]{Alexander2020}. Radio properties of TDEs are also highly diverse, with some events showing prompt emission consistent with being ejected very early during the debris circularisation \citep[e.g.][]{Cendes2021,Goodwin2022,Goodwin2023,Goodwin2024} and others showing late rising radio emission thought to be due to delayed jets or outflows, new interactions of a pre-existing outflow with the CNM, or an off-axis jet coming into view \citep[e.g.][]{Cendes2022,Horesh2021,Horesh2021b,Hajela2024,Matsumoto2023}.

The diversity in TDE observational properties at all wavelengths has been argued to be due to varying physical parameters of the disruption, such as stellar mass, SMBH mass, SMBH spin, stellar orbit and structure, impact parameter, host galaxy environment, SMBH accretion history, and the viewing angle \citep[e.g.][]{Lodato2009,Guillochon2015,Dai2018}. In the scenario where the visibility of optical or X-ray emission is due to differences in viewing angles, TDEs selected at X-ray wavelengths may probe events in which the optical flare is obscured by the fast outflows or jet  \citep{Dai2018}. 

In other accreting black hole systems, such as X-ray binaries, a positive correlation between radio and X-ray luminosity is well-established in the low-luminosity hard X-ray state \citep[e.g.][]{Gallo2003,Brinkmann2000}. With the addition of a mass term, this correlation can also be extended to AGN, forming the Fundamental Plane of Black Hole Activity \citep{Merloni2003,Falcke2004}. It is thought that this correlation is driven by a correlation between jet power and mass accretion rate during the hard state \citep[e.g.][]{Falcke2004}. However, this correlation has not been seen for the transient radio-emitting ejecta that are released during the hard-to-soft state transition in X-ray binaries \citep{Fender2004}. In the case of a TDE, the presence of a correlation between X-ray and radio emission has not been extensively studied. Long-term UV monitoring of TDEs has revealed evidence of long-lived accretion disks in these systems \citep{Mummery2024}, and \citet{Piro2024} suggested radio outflows may be ejected hundreds of days after a TDE due to thermal instabilities in the long-lived accretion disk. \citet{Sfaradi2022} claimed a tentative correlation between a radio flare associated with an X-ray state change in the TDE AT2019azh, but \citet{Goodwin2022} showed that the bumpy radio lightcurve of AT2019azh could be due to an inhomogenous circumnuclear medium without requiring an energy injection episode. Radio observations of an X-ray selected sample of TDEs are required in order to truly constrain any relationship between radio and X-ray emission in these events.

To date, studies of samples of TDEs in the radio have been limited to optically-discovered events \citep[e.g.][]{Cendes2024}, which may not encapsulate the full observational distribution of TDEs. Observing a systematically X-ray selected sample of events in the radio is necessary to understand the conditions under which detectable radio emission is produced and to search for any correlation with X-ray emission.

In this work we present a systematic radio follow-up campaign of 22 TDE candidates discovered by eROSITA during its first and second all-sky scans in 2019--2020. In Section \ref{sec:sample} we present the sample selection criteria and candidates in the sample, in Section \ref{sec:radio_obs} we present details of the radio observations of each candidate in the sample, in Section \ref{sec:results} we present the radio lightcurves, spectra, fits, and outflow models of the detected events, in Section \ref{sec:discussion} we discuss the nature of the radio emission of the detected sources, compare to radio outflows observed from optically-selected TDEs, and discuss the implications of these results for the outflow mechanism. Finally, in Section \ref{sec:summary} we summarise the results and provide concluding remarks.

\section{Sample Selection}\label{sec:sample}

The eROSITA sample of TDE candidates we followed up was identified through systematically searching eRO-ExTra, a catalog of non-AGN extragalactic transients and variables from eRASS1 and eRASS2 
\citep{Grotova2025}, for highly variable sources that:
\begin{itemize}
    \item were at their peak observed 0.2--2\,keV flux in the eRASS1 or eRASS2 epochs and only showed a single decay or flare-like X-ray lightcurve
    \item showed large amplitude X-ray flares ($\gtrsim4\times$ change in 0.2--2\,keV flux between eRASS epochs)
    \item showed ultra-soft X-ray emission (photon indices $>3$)
    \item were associated with a quiescent host galaxy, classified as non-active on the basis of: i) its non-AGN like mid-infrared colour (WISE W1-W2$<$0.8), ii) there being no archival X-ray detection of the system (e.g. with ROSAT, XMM), iii) its optical spectrum showing no strong evidence for AGN activity
\end{itemize}

These conservative selection criteria for very strong TDE candidates discovered by eROSITA resulted in 31 sources. The sample and optical spectroscopy of the sources are presented in \textcolor{black}{\citet{Grotova_inpress}}. 

In this work, we sub-selected the TDE candidates for sources that are visible to the Australia Telescope Compact Array (ATCA; Declination $<15$\,deg  and not $-4<$ Dec$<4$\,deg), that had a redshift measurement available at the time the sample selection was carried out, and were associated with a nearby host galaxy at $z<0.35$. This selection resulted in 22 TDE candidates in our sample for which we obtained dedicated ATCA radio follow-up observations. The coordinates and redshifts of each of the 22 TDEs in our sample are listed in Table \ref{tab:sample}. The majority of the TDEs have spectrometric redshift measurements, however 5 TDEs in the sample only have photometric redshift estimates available. See \textcolor{black}{\citet{Grotova_inpress}} for a discussion on how the redshift measurements were calculated. 
The X-ray flux measurements, date of peak eROSITA X-ray flux, and X-ray spectral properties of each source are presented in \textcolor{black}{\citet{Grotova_inpress}}. 

\begin{table}[!ht]
    \centering
    \caption{The eROSITA TDE candidates in our sample}\label{tab:sample}
    \begin{tabular}{llll}
    \hline
        Source & RA & DEC & z \\ 
        \hline
       eRASSt J011431-593654 & 18.6285924 & -59.615177 & 0.16 \\ 
       eRASSt J014133-443415 & 25.3891026 & -44.570699 & 0.092 \\ 
       eRASSt J031857-205452 & 49.7387829 & -20.914328 & 0.27$_{-0.09}^{+0.13*}$ \\ 
       eRASSt J043959.5-651403 & 69.9985529 & -65.234207 & 0.15 \\ 
       eRASSt J060829-435319 & 92.1225777 & -43.889217 & 0.076 \\ 
       eRASSt J063413-713908 & 98.5557613 & -71.651827 & 0.13 \\ 
       eRASSt J064449-603704 & 101.205218 & -60.61777 & 0.12 \\ 
       eRASSt J070710-842044 & 106.789457 & -84.346016 & 0.17$_{-0.02}^{+0.23*}$ \\ 
       eRASSt J082056+192538 & 125.232368 & 19.4269809 & 0.12 \\ 
       eRASSt J082337+042302 & 125.903203 & 4.38401612 & 0.028 \\ 
       eRASSt J091658+060956 & 139.240653 & 6.16582279 & 0.058 \\ 
       eRASSt J110240-051813 & 165.66726 & -5.3030853 & 0.33$_{-0.09}^{+0.10*}$ \\ 
       eRASSt J140158+081402 & 210.492026 & 8.23370706 & 0.26$_{-0.05}^{+0.06*}$ \\ 
       eRASSt J142140-295321 & 215.415504 & -29.889602 & 0.056 \\ 
       eRASSt J143624-174105 & 219.098949 & -17.684542 & 0.19$_{-0.03}^{+0.06*}$ \\ 
       eRASSt J145623-283853 & 224.095823 & -28.648657 & 0.05 \\ 
       eRASSt J164649-692539 & 251.703227 & -69.427139 & 0.017 \\ 
       eRASSt J190147-552200 & 285.446108 & -55.36617 & 0.058 \\ 
       eRASSt J210858-562832 & 317.242446 & -56.475326 & 0.043 \\ 
       eRASSt J153406-090332 & 233.526108 & -9.0592701 & 0.023 \\ 
       eRASSt J234403-352641 & 356.012306 & -35.444909 & 0.1 \\
       eRASSt J145954.5-260822 & 224.978036 & -26.140874 & 0.09 \\
        \hline
    \end{tabular}
    $^*$photometric redshift, see \citet{Grotova2025} and \textcolor{black}{\citet{Grotova_inpress}} for a discussion on how this was calculated.
\end{table}

\section{Radio Observations}\label{sec:radio_obs}

\subsection{ATCA observations}

In order to search for transient radio emission associated with the X-ray flares, we observed the coordinates of each TDE candidate in the sample a minimum of two times with ATCA. The two observations of each source were spaced approximately 6 months apart in order to ascertain if detected sources were radio variable or consistent with non-variable host emission. 5 sources were observed between October 2022 and June 2023 (ATCA program C3513; PI Goodwin) and 15 sources were observed between April and October 2024 (ATCA program C3513; PI Goodwin). 

The remaining two sources in the sample (eRASSt J234403-352641 and eRASSt J164649-692539) were observed via a separate triggered ATCA program (program C3334; PI: Goodwin) in an independent study. eRASSt J234403-352641 has variable radio emission presented in detail by \citet{Goodwin2024}, with numerous epochs of ATCA observations taken over 2021--2024 at frequencies between 1--15\,GHz. To compare the radio observations of this source with the rest of the sample, we selected radio observations approximately 6 months apart at 2.1, 5.5, 9\,GHz, at a similar time post X-ray detection to the observations in the rest of the sample. The radio data reduction and flux density extraction process are outlined in \citet{Goodwin2024}. eRASSt J164649-692539 was observed twice: at central frequencies of 5.5, and 9\,GHz in August 2022 and at central frequencies of 2.1, 5.5, and 9\,GHz in July 2024, and the data reduced as described below.

All sources were observed using the 4\,cm observing band with the dual 5.5 and 9\,GHz receivers with 2\,GHz of bandwidth at each frequency divided into 2048 spectral channels. Additionally, some sources were observed in the 2.1\,GHz observing band with 2\,GHz of bandwidth divided into 2048 spectral channels. Observations were carried out with ATCA in the most extended 6\,km configuration in order to maximise the resolution of the images and minimise the possibility of source confusion. At 2.1\,GHz this configuration produces a synthesised beam of 3-9", at 5.5 and 9\,GHz of 1-3", and at 17 and 21\,GHz of 0.5". 

All data were reduced in the Common Astronomy Software Application \citep[CASA v 5.6;][]{CASA2022} using standard procedures including flux and bandpass calibration with 1934--638 or 0823--500 (when 1934--638 was not visible). Secondary calibration was carried out with a calibrator source for each target. Each source was observed for a total of 4\,hr for each observation alternating between 15--20\,min target scans and 2\,min secondary calibrator scans. Images of each target field were made using the CASA task \texttt{tclean}. When a source was detected at the coordinates of the target, a flux density was extracted using the CASA task \texttt{imfit} and fitting a Gaussian the size of the synthesised beam. Flux density errors were calculated by adding in quadrature the statistical error and a 5$\%$ additional systematic error to account for the absolute accuracy of the flux density scale. When no source was detected at the coordinates of the target a 3$\sigma$ upper limit was measured as 3 times the image rms.

A summary of the radio observations including flux densities and upper limits measured for each epoch of each source is given in Table \ref{tab:radio_obs}.

\begin{longtable}{p{2.5cm}lllll}
\caption{ATCA radio observations of each of the 22 TDE candidates in our sample}\\
\hline
Source & Epoch Date & $\Delta t$ (d)$^*$ & $\nu$\,(GHz) & $F_{\nu}$\,($\rm{\mu}$Jy)\\
\hline
eRASSt & 2024-04-01 & 1224 & 5.0 & 160$\pm$26\\
J011431-593654 & 2024-04-01 & 1224 & 5.5 & 175$\pm$19\\
& 2024-04-01 & 1224 & 6.0 & 192$\pm$24\\
& 2024-04-01 & 1224 & 8.5 & 139$\pm$21\\
& 2024-04-01 & 1224 & 9.0 & 123$\pm$17\\
& 2024-04-01 & 1224 & 9.5 & 115$\pm$20\\
& 2024-10-09 & 1415 & 5.0 & 191$\pm$16\\
& 2024-10-09 & 1415 & 5.5 & 183$\pm$12\\
& 2024-10-09 & 1415 & 6.0 & 160$\pm$15\\
& 2024-10-09 & 1415 & 8.5 & 184$\pm$14\\
& 2024-10-09 & 1415 & 9.0 & 187$\pm$9\\
& 2024-10-09 & 1415 & 9.5 & 201$\pm$11\\
\hline
eRASSt & 2024-04-02 & 1482 & 5.5 & $<$63\\
J014133-443415& 2024-04-02 & 1482 & 9.0 & $<$69\\
& 2024-10-09 & 1672 & 5.5 & $<$42\\
& 2024-10-09 & 1672 & 9.0 & $<$34\\
\hline
eRASSt & 2024-04-01 & 1341 & 5.5 & $<$45\\
J031857-205452& 2024-04-01 & 1341 & 9.0 & $<$42\\
& 2024-10-09 & 1532 & 5.5 & 80$\pm$5\\
& 2024-10-09 & 1532 & 9.0 & $<$34\\
\hline
eRASSt & 2022-10-02 & 934 & 5.5 & 53$\pm$10\\
J043959.5-651403& 2022-10-02 & 934 & 9.0 & $<$39\\
& 2023-06-17 & 1192 & 5.5 & 61$\pm$17\\
& 2023-06-17 & 1192 & 9.0 & 39$\pm$9\\
\hline
eRASSt & 2024-04-02 & 1299 & 5.5 & $<$51\\
J060829-435319 & 2024-04-02 & 1299 & 9.0 & $<$48\\
& 2024-10-14 & 1493 & 5.5 & $<$42\\
& 2024-10-14 & 1493 & 9.0 & $<$45\\
\hline
eRASSt & 2024-04-02 & 1430 & 5.5 & 93$\pm$28\\
J063413-713908 & 2024-04-02 & 1430 & 9.0 & $<$93\\
& 2024-10-13 & 1623 & 5.5 & 79$\pm$19\\
& 2024-10-13 & 1623 & 9.0 & $<$55\\
\hline
eRASSt & 2024-04-01 & 1280 & 5.5 & $<$45\\
J064449-603704 & 2024-04-01 & 1280 & 9.0 & $<$42\\
& 2024-10-11 & 1476 & 5.5 & $<$37\\
& 2024-10-11 & 1476 & 9.0 & $<$29\\
\hline
eRASSt & 2024-04-01 & 1285 & 5.5 & $<$60\\
J070710-842044 & 2024-04-01 & 1285 & 9.0 & $<$54\\
& 2024-10-14 & 1480 & 5.5 & $<$56\\
& 2024-10-14 & 1480 & 9.0 & $<$70\\
\hline
eRASSt & 2024-04-02 & 1251 & 5.5 & $<$48\\
J082056+192538 & 2024-04-02 & 1251 & 9.0 & $<$42\\
& 2024-10-14 & 1446 & 5.5 & $<$45\\
& 2024-10-14 & 1446 & 9.0 & $<$33\\
\hline
eRASSt & 2024-04-03 & 1251 & 5.0 & 1047$\pm$59\\
J082337+042302 & 2024-04-03 & 1251 & 5.5 & 989$\pm$52\\
& 2024-04-03 & 1251 & 6.0 & 982$\pm$62\\
& 2024-04-03 & 1251 & 8.5 & 772$\pm$45\\
& 2024-04-03 & 1251 & 9.0 & 771$\pm$40\\
& 2024-04-03 & 1251 & 9.5 & 712$\pm$43\\
& 2024-10-15 & 1446 & 5.0 & 791$\pm$24\\
& 2024-10-15 & 1446 & 5.5 & 756$\pm$16\\
& 2024-10-15 & 1446 & 6.0 & 735$\pm$20\\
& 2024-10-15 & 1446 & 8.5 & 652$\pm$25\\
& 2024-10-15 & 1446 & 9.0 & 594$\pm$16\\
& 2024-10-15 & 1446 & 9.5 & 536$\pm$18\\
\hline
eRASSt & 2022-10-02 & 877 & 5.5 & $<$78\\
J091658+060956& 2022-10-02 & 877 & 9.0 & $<$42\\
& 2023-06-12 & 1130 & 5.5 & $<$63\\
& 2023-06-12 & 1130 & 9.0 & $<$60\\
\hline
eRASSt & 2024-04-03 & 1369 & 5.5 & 127$\pm$16\\
J110240-051813 & 2024-04-03 & 1369 & 9.0 & 95$\pm$12\\
& 2024-10-15 & 1562 & 5.5 & 120$\pm$15\\
& 2024-10-15 & 1562 & 9.0 & 82$\pm$13\\
\hline
eRASSt & 2024-04-06 & 1340 & 5.5 & $<$42\\
J140158+081402 & 2024-04-06 & 1340 & 9.0 & $<$36\\
& 2024-10-16 & 1533 & 5.5 & $<$59\\
& 2024-10-16 & 1533 & 9.0 & 83$\pm$17\\
\hline
eRASSt & 2024-04-04 & 1526 & 5.5 & 85$\pm$19\\
J142140-295321 & 2024-04-04 & 1526 & 9.0 & 64$\pm$19\\
& 2024-10-14 & 1720 & 5.5 & $<$60\\
& 2024-10-14 & 1720 & 9.0 & $<$186\\
\hline
eRASSt & 2024-04-05 & 777 & 5.5 & 72$\pm$21\\
J143624-174105 & 2024-04-05 & 777 & 9.0 & $<$75\\
& 2024-10-16 & 1033 & 5.5 & $<$45\\
& 2024-10-16 & 1033 & 9.0 & $<$59\\
\hline
eRASSt & 2022-10-02 & 1326 & 5.5 & $<$52\\
J145623-283853 & 2022-10-02 & 1326 & 9.0 & $<$42\\
& 2023-06-15 & 1522 & 5.5 & $<$34\\
& 2023-06-15 & 1522 & 9.0 & $<$29\\
\hline
eRASSt & 2024-04-02 & 954 & 5.5 & $<$42\\
J145954.5-260822 & 2024-04-02 & 954 & 9.0 & 47$\pm$13\\
& 2024-10-15 & 1205 & 5.5 & $<$46\\
& 2024-10-15 & 1205 & 9.0 & $<$33\\
\hline
eRASSt & 2022-10-03 & 702 & 5.5 & $<$43.00\\
J153406-090332 & 2022-10-03 & 702 & 9.0 & $<$35\\
& 2023-06-11 & 1387 & 5.5 & $<$31\\
& 2023-06-11 & 1387 & 9.0 & $<$30\\
\hline
eRASSt & 2022-08-28 & 668 & 5.0 & 183$\pm$11\\
J164649-692539 & 2022-08-28 & 668 & 5.5 & 147$\pm$11\\
& 2022-08-28 & 668 & 6.0 & 136$\pm$14\\
& 2022-08-28 & 668 & 8.5 & 74$\pm$12\\
& 2022-08-28 & 668 & 9.0 & 70$\pm$9\\
& 2022-08-28 & 668 & 9.5 & 66$\pm$11\\
& 2024-07-13 & 1353 & 2.1 & 604$\pm$31\\
 & 2024-07-13 & 1353 & 5.0 & 162$\pm$23\\
& 2024-07-13 & 1353 & 5.5 & 141$\pm$16\\
& 2024-07-13 & 1353 & 6.0 & 128$\pm$21\\
& 2024-07-13 & 1353 & 8.5 & 105$\pm$19\\
& 2024-07-13 & 1353 & 9.0 & 105$\pm$14\\
& 2024-07-13 & 1353 & 9.5 & 89$\pm$16\\
\hline
eRASSt & 2022-10-03 & 908 & 5.0 & 374$\pm$25\\
J190147-552200 & 2022-10-03 & 908 & 5.5 & 352$\pm$12\\
& 2022-10-03 & 908 & 6.0 & 311$\pm$22\\
& 2022-10-03 & 908 & 8.5 & 180$\pm$16\\
& 2022-10-03 & 908 & 9.0 & 167$\pm$10\\
& 2022-10-03 & 908 & 9.5 & 155$\pm$12\\
& 2023-06-30 & 1178 & 2.1 & 590$\pm$32\\
 & 2023-06-30 & 1178 & 5.0 & 322$\pm$13\\
& 2023-06-30 & 1178 & 5.5 & 300$\pm$13\\
& 2023-06-30 & 1178 & 6.0 & 281$\pm$19\\
& 2023-06-30 & 1178 & 8.5 & 149$\pm$21\\
& 2023-06-30 & 1178 & 9.0 & 116$\pm$16\\
& 2023-06-30 & 1178 & 9.5 & 116$\pm$16\\
\hline
eRASSt & 2024-04-05 & 1257 & 5.0 & 456$\pm$37\\
J210858-562832 & 2024-04-05 & 1257 & 5.5 & 460$\pm$28\\
& 2024-04-05 & 1257 & 6.0 & 391$\pm$39\\
& 2024-04-05 & 1257 & 8.5 & 132$\pm$33\\
& 2024-04-05 & 1257 & 9.0 & 263$\pm$22\\
& 2024-04-05 & 1257 & 9.5 & 164$\pm$38\\
& 2024-10-14 & 1449 & 5.0 & 313$\pm$18\\
& 2024-10-14 & 1449 & 5.5 & 291$\pm$12\\
& 2024-10-14 & 1449 & 6.0 & 277$\pm$16\\
& 2024-10-14 & 1449 & 8.5 & 223$\pm$15\\
& 2024-10-14 & 1449 & 9.0 & 199$\pm$10\\
& 2024-10-14 & 1449 & 9.5 & 179$\pm$14\\
\hline
eRASSt & 2022-12-11 & 746 & 5.0 & 649$\pm$25\\
J234403-352641 & 2022-12-11 & 746 & 5.5 & 649$\pm$25\\
& 2022-12-11 & 746 & 6.0 & 609$\pm$18\\
& 2022-12-11 & 746 & 8.5 & 467$\pm$16\\
& 2022-12-11 & 746 & 9.0 & 399$\pm$18\\
& 2022-12-11 & 746 & 9.5 & 399$\pm$18\\
& 2023-06-07 & 924 & 2.1 & 316$\pm$44\\
 & 2023-06-07 & 924 & 5.0 & 611$\pm$22\\
& 2023-06-07 & 924 & 5.5 & 611$\pm$22\\
& 2023-06-07 & 924 & 6.0 & 616$\pm$20\\
& 2023-06-07 & 924 & 8.5 & 395$\pm$19\\
& 2023-06-07 & 924 & 9.0 & 303$\pm$18\\
& 2023-06-07 & 924 & 9.5 & 303$\pm$18 \\
\hline
\label{tab:radio_obs}
\footnotesize{$^*$$\Delta t$ is measured assuming $t_0$ is the time of the peak eROSITA X-ray flux measurement.}
\end{longtable}

\subsection{Archival radio observations}
In addition to the dedicated ATCA observations, we also searched recent public archival radio survey data. Whilst these observations are significantly less sensitive than dedicated follow-up observations, they enable us to obtain constraints on any pre-existing radio emission or new transient radio emission at lower frequencies from each source. We examined observations taken by the Very Large Array Sky Survey \citep[VLASS;][]{VLASS_paper} at a central frequency of 3\,GHz and the Australian Square Kilometre Array Pathfinder (ASKAP) Rapid ASKAP Continuum Survey \citep[RACS][]{RACs1,RACs2,RACs3} at a central frequency of 0.9 and 1.4\,GHz. 

\subsubsection{RACS}
ASKAP-RACS observations were available for all 22 sources in the sample, with multiple epochs taken between 2019 and 2024. We downloaded image cutouts centered on each source using the CASDA online portal. We selected publicly available RACS epoch observations instead of the combined RACS data release images in order to retain temporal information about each epoch of observations. The majority of sources had two RACS observations available at 0.9\,GHz and one or two RACS observations available at 1.4\,GHz. For each image cutout, we measured the peak brightness in a 0.5' aperture at the location of each source, and compared to the background rms of the image. We visually examined images of all sources with a peak flux greater than 3 times the rms, and found 3 sources with significant detections. We report the measured flux densities and 3$\sigma$ upper limits for each source in the Appendix in Table \ref{tab:RACS_data}. 

\subsubsection{VLASS}
VLASS observations were available for 12 out of 22 sources in the sample, with multiple epochs taken between 2017 and 2024. We downloaded image cutouts centered on each source using the CIRADA online portal. The majority of sources had three VLASS observations available at 3\,GHz with epochs spaced approximately 2\,yr apart between 2017--2024. We measured the peak brightness in a 0.5' aperture at the location of each source, and compared to the background rms of the image. We visually examined images of all sources with a peak flux greater than 3 times the rms, and found 1 source with a significant detection. We report the measured flux densities and 3$\sigma$ upper limits for each source in the Appendix in Table \ref{tab:VLASS_data}. 

\section{Results}\label{sec:results}
\begin{figure}
    \centering
    \includegraphics[width=1.1\linewidth]{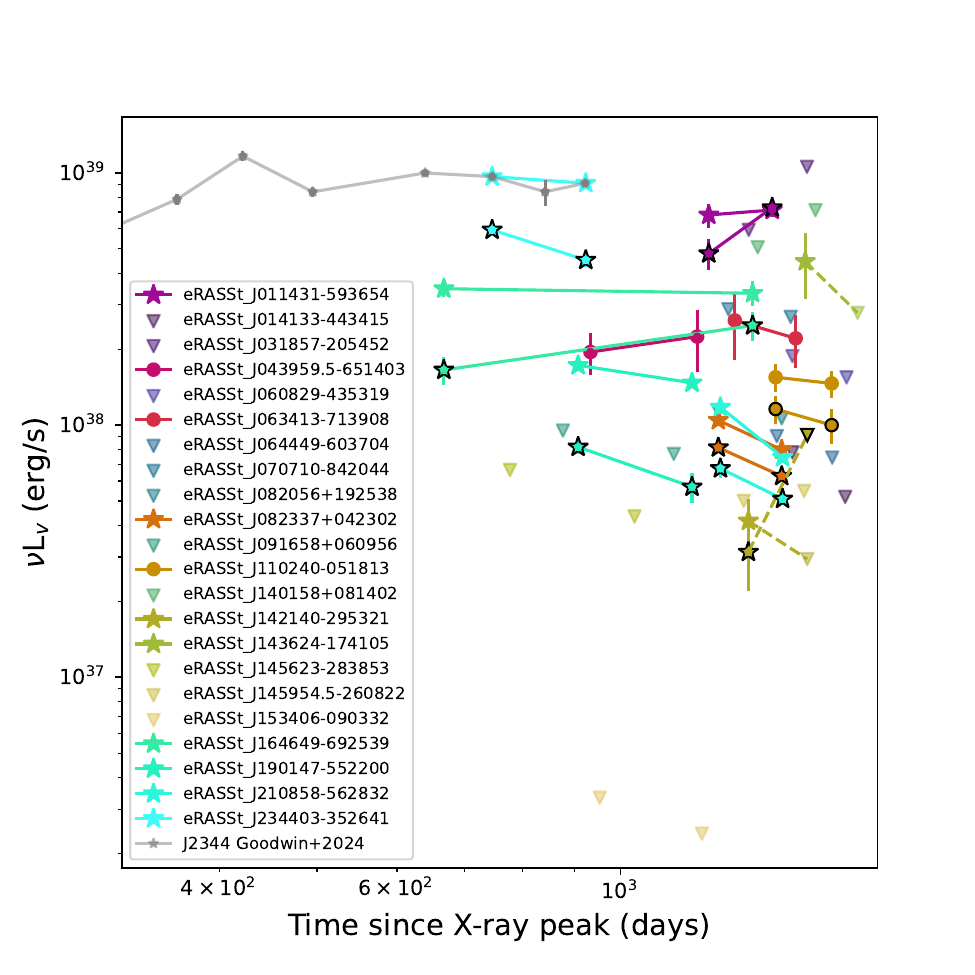}
    \caption{The 5.5\,GHz (no outline) and 9\,GHz (black outline) radio lightcurves between the two ATCA epochs of the 22 TDEs in our sample. The sources with statistically significant radio variability are shown as stars, the sources with no statistically significant radio variability are shown in circles, and all non-detections are the 5.5\,GHz 3$\sigma$ upper limits plotted as inverted triangles. The grey stars show additional 5.5\,GHz radio data of eRASSt J234403-352641 from \citet{Goodwin2024}. Of the 22 sources in the sample, 11 were radio-detected, and 8 showed statistically significant radio variability at either 5.5 or 9\,GHz. }
    \label{fig:rad_lcs}
\end{figure}

Our radio observations reveal radio sources coincident with half of the TDEs in the sample (11/22). The observed 5.5\,GHz radio lightcurves and upper limits of each of the sources in the sample are plotted in Figure \ref{fig:rad_lcs}. Overall the radio luminosities range from $10^{37}$--$10^{39}$\,erg/s at times of 600--1500\,d post X-ray peak luminosity. 

\subsection{Radio variability properties}
In order to determine the significance of any variability detected between the two ATCA epochs, we calculate two variability parameters. The first is the variability index, V, at 5.5 and 9\,GHz for each source as \citep[e.g.][]{Hovatta2008}
\begin{equation}\label{eq:var}
    V = \frac{(S_{\rm{max}} - \sigma_{S\rm{max}}) - (S_{\rm{min}} + \sigma_{S\rm{min}})}{(S_{\rm{max}} - \sigma_{S\rm{max}}) +(S_{\rm{min}} + \sigma_{S\rm{min}})},
\end{equation}
where $S_{\rm{max}}$ and $S_{\rm{min}}$ are the maximum and minimum flux densities measured for each source, and $\sigma_{S\rm{max}}$ and $\sigma_{S\rm{min}}$ their respective errors. This parameter measures whether any variability detected is statistically significant. Any value of $V>0$ indicates statistically significant variability between the two epochs of observations. 

Secondly, we calculate a modulation fraction $m_f$ of the variability detected for each source, as

\begin{equation}\label{eq:mf}
    m_{f,obs} = \frac{S_{\rm{1}}}{S_{\rm{2}}} - 1
\end{equation}
where $S_{\rm{1}}$ is the flux density of the first epoch and $S_{\rm{2}}$ is the flux density of the second epoch. 

The modulation fraction does not account for errors on the observed flux densities, but enables comparison between the magnitude of variability observed and the expected magnitude of variability from interstellar scintillation.

The variability indices and modulation fractions for each of the 11 detected sources are given in Table \ref{tab:Vpars}. We detected statistically significant radio variability at one or both observing frequencies in 8 sources. 

Including the archival RACS and VLASS observations, we note that eRASSt J082337+042302 shows statistically significant radio variability at 3\,GHz in VLASS observations (as well as in our ATCA observations), but unfortunately there was only one epoch of RACS observations at 0.9 and 1.4\,GHz so there are no variability constraints at those frequencies. eRASSt J164649-692539 and eRASSt J210858-562832 both were observed multiple times with RACS between 2019 and 2024 but did not show statistically significant variability at 0.9 or 1.4\,GHz. 

\begin{table*}[]
   \centering
\caption{Variability statistics for the radio-detected sources between the two ATCA epochs at 5.5 and 9\,GHz and the calculated ISS parameters for the Galactic location of each source calculated assuming a conservative source size of $10^{16.5}$\,cm.}
\begin{tblr}{colspec={X[l] X[l] X[l] X[l] X[l] X[l] X[l] X[l] X[l] X[l] X[l] X[l] X[l]}}%
\hline
Source & $V$  & $m_{f,obs}$ &$V$ & $m_{f,obs}$ & $\nu_F$ & $\theta_F$ & $m_{f,ISS}$ & $t_{f,ISS}$ & $m_ {f,ISS}$  & $t_{f,ISS}$  & Sig. & $m_{f,obs}>$\\
 &5.5GHz & 5.5GHz &9\,GHz & 9GHz & (GHz) &  &  5.5GHz & 5.5\,GHz (hr) & 9\,GHz & 9\,GHz (hr) & V? & $m_{f,ISS}$? \\
\hline
eRASSt J011431-593654 & -0.07 & 0.05 & 0.13 & 0.52&8.26 & 0.79 & 0.52 & 7.01 & 0.07 & 16.18 & Yes & Yes \\
eRASSt J043959.5-651403 & -0.16 & 0.15 & - & - & 9.87 & 0.88 & 0.72 & 7.24 & 0.16 & 11.17 &No &No \\
eRASSt J063413-713908 & -0.18 & 0.18 & - & -& 12.39 & 1.12 & 0.63 & 11.93 & 0.46 & 6.72 & No &No \\
eRASSt J082337+042302 & 0.09 & 0.31 & 0.09 & 0.30 & 12.59 & 1.05 & 0.48 & 15.52 & 0.07 & 35.86 & Yes & Yes \\
eRASSt J110240-051813 & -0.10 & 0.06 & -0.07 & 0.16& 9.00 & 0.91 & 0.76 & 5.91 & 0.22 & 7.28 & No & No \\
eRASSt J142140-295321 & $>$0.05 & $>$0.42 & $>$-0.6 & $>$1.91& 12.97 & 1.34 & 0.61 & 13.21 & 0.23 & 13.12 & Yes & Yes \\
eRASSt J143624-174105 & $>$0.06 & $>$0.60 & - & -& 10.40 & 0.98 & 0.31 & 16.43 & 0.04 & 37.95 & Yes & Yes \\
eRASSt J164649-692539 & -0.07 & 0.04 & 0.07 & 0.50& 23.01 & 2.88 & 0.44 & 46.62 & 0.59 & 15.78 & Yes & No \\
eRASSt J190147-552200 & 0.04 & 0.17 & 0.09 & 0.44& 15.90 & 1.78 & 0.55 & 20.67 & 0.72 & 6.99 & Yes & No \\
eRASSt J210858-562832 & 0.17 & 0.58 & 0.07 & 0.32& 10.30 & 1.06 & 0.36 & 13.98 & 0.05 & 32.30 & Yes & Yes \\
eRASSt J234403-352641 & -0.01 & 0.06 & 0.09 & 0.32& 7.75 & 0.78 & 0.23 & 12.65 & 0.03 & 29.23 & Yes & Yes \\
\hline
\end{tblr}
\footnotesize{$V$ is the variability index (calculated using Equation \ref{eq:var}), $m_{f,obs}$ is the observed modulation fraction (calculated using Equation \ref{eq:mf}), $\nu_F$ is the ISS transition frequency, $\theta_F$ is the angular size of the First Fresnel zone, and $m_{f,ISS}$ is the expected modulation fraction on timescale $t_{f,ISS}$ due to ISS. }
    \label{tab:Vpars}
\end{table*}

\subsubsection{Interstellar scintillation}
After establishing that we detected statistically significant radio variability of 8 sources, we next need to assess whether this radio variability is intrinsic to the radio source, or consistent with extrinsic variability due to interstellar scintillation (ISS). 
ISS can induce radio variability due to the scattering of radio waves as they pass through gas in the Galaxy. For compact extra-galactic radio sources such as those in our sample, the magnitude and timescale of radio variability due to ISS is dependent on the Galactic coordinates of the target, and how much gas is along the line of sight to the source \citep{Walker1998}. In order to assess if any variability we observed may be due to ISS rather than intrinsic source variability, we calculated the expected modulation fraction and timescale of modulation for each of the radio-detected sources in our sample. We used the NE2001 Galactic electron density model \citep{Cordes2002} to determine a transition frequency ($\nu_F$) and angular size limit of the first Fresnel zone ($\theta_f$) at the Galactic coordinates of each source. Using the \citet{Walker1998} formalism, we infer that the radio emission from most sources will be in either the strong or weak refractive scintillation regimes, assuming a conservative source size of $10^{16.5}$\,cm at observing frequencies of 5.5 and 9\,GHz. We find for our sources the variability expected at these frequencies has modulation fractions ranging from 0.2--0.8 at 5.5\,GHz on timescales of 5--47\,hr and  modulation fractions ranging from 0.03--0.7 at 9\,GHz on timescales of 6--38\,hr. In Table \ref{tab:Vpars} we list the calculated transition frequencies, angular size limits of the first Fresnel zone, scintillation modulation fractions ($m_{f,ISS}$), and timescale of modulation ($t_{f,ISS}$) for each of the radio-detected sources at each central observing frequency. The reported modulation fractions are heavily dependent on the NE2001 Galactic electron density model as well as the assumed source size, and may change significantly with different assumed Galactic electron density distributions and source sizes. The values reported in the table should be used as a guide only for the expected magnitude of variability induced by ISS.  

We find that the observed 5.5\,GHz variability modulation fractions (reported in Table \ref{tab:Vpars}) for all sources except one (eRASSt J143624-174105) are consistent with the variability expected due to ISS. Whereas, at 9\,GHz we find that the observed 9\,GHz variability modulation fractions for each source are larger than expected due to ISS except for eRASSt J164649-692539 and eRASSt J190147-552200. We therefore deduce that of the 8 sources with statistically significant radio variability, 6 of the sources showed intrinsic radio variability greater than expected if it were due only to ISS. 

\subsection{Radio Spectral Fitting}\label{sec:specfits}

In order to understand the nature of the radio emission detected, we carried out spectral fitting with the goal of assessing whether or not the radio spectra of each detected source are consistent with a synchrotron self absorption model, as expected for a newly launched outflow from a TDE. We fit the spectra of each epoch of the 8 sources detected at both ATCA observing frequencies of 5.5 and 9\,GHz using the following equation

\begin{equation}
\label{eq:specindex}
\alpha=\frac{\log{F_1 / F_2}}{\log{ \nu_1 / \nu_2}} 
\end{equation}
where $\alpha$ is the spectral index, and $F$ the flux density measured at frequency $\nu$. Error on $\alpha$ was calculated using standard error propagation of the flux density errors.

The spectral indices for each epoch of each source are given in Table \ref{tab:specindices}. Most sources show spectral indices consistent with optically thin synchrotron emission \citep[e.g.][]{Granot2002}, although the spectra of eRASSt J210858-562832 and eRASSt J190147-552200 are much steeper in the second epoch, implying either a synchrotron spectral index, $p>3$ or a very old/large synchrotron-emitting region. 

\begin{table}[]
    \centering
    \caption{Measured radio spectral indices between 5.5 and 9\,GHz for sources detected in both ATCA observing bands.}
    \begin{tabular}{ccc}
\hline
Source & $\alpha$ epoch 1 & $\alpha$ epoch 2\\
\hline
eRASSt J011431-593654 & -0.72$\pm$0.03 & 0.040$\pm$0.006\\
eRASSt J043959.5-651403  & undetected & -0.9$\pm$0.1\\
eRASSt J082337+042302 & -0.510$\pm$0.005 & -0.490$\pm$0.001\\
eRASSt J110240-051813 & -0.59$\pm$0.03 & -0.77$\pm$0.04\\
eRASSt J142140-295321 & -0.58$\pm$0.12 & undetected\\
eRASSt J164649-692539 & -1.51$\pm$0.02 & -0.60$\pm$0.03\\
eRASSt J190147-552200 & -1.510$\pm$0.004 & -1.93$\pm$0.014\\
eRASSt J210858-562832 & -1.140$\pm$0.009 & -0.770$\pm$0.004\\
eRASSt J234403-352641 & -0.990$\pm$0.003 & -1.420$\pm$0.004\\
\hline
    \end{tabular}

    \label{tab:specindices}
\end{table}

In addition to the two-point spectral fits, for the 6 sources that were bright enough to split into sub-bands with ATCA we additionally fit a physical model for synchrotron self-absorption for each of the two epochs. We use the \citet{Granot2002} model, as is standard for TDE radio spectral modelling \citep[e.g.][]{Alexander2016,Goodwin2022,Cendes2021}, in which the peak of the synchrotron spectrum is associated with the self-absorption break ($\nu_a$), i.e. $\nu_{\rm m} < \nu_{\rm a} < \nu_{\rm c}$,  where $\nu_{\rm c}$ is the synchrotron cooling frequency. In this model, the flux density of the self-absorbed synchrotron component is given by
\begin{equation}
    \label{eq:Fnua}
        F_{\nu, \mathrm{synch}} = F_{\nu,\mathrm{ext}} \left[
        \left(\frac{\nu}{\nu_{\rm a}}\right)^{-s\alpha_1} +  \left(
        \frac{\nu}{\nu_{\rm a}}\right)^{-s\alpha_2)
        }\right]^{-1/s}
    \end{equation}
    where $\nu$ is the frequency, $F_{\nu,\mathrm{ext}}$ is the normalisation, $s = 1.25-0.18p$ (which assumes a wind-like density profile with $k=2$), $\alpha_1 = \frac{5}{2}$, $\alpha_2 = \frac{1-p}{2}$, assuming a single break is observed in the spectrum and the peak is associated with the self-absorption break. 

We fit the spectra using a Python implementation of Markov Chain Monte Carlo (MCMC), \texttt{emcee} \citep{Mackey2013}. We allow the model parameters to vary inside flat in log-space prior distributions, where we constrain $p$ to the range 2.0--4.0, $F_{\nu,\rm{ext}}$ to the range $10^{-6}$--$10$\,mJy, and $\nu_a$ to the range 0.1--9\,GHz.  For each parameter we report the median value from the posterior distribution and the 16th and 84th percentiles, corresponding to approximately 1$\sigma$ errors. For each MCMC calculation we use 400 walkers and 5000 steps, discarding the first 500 steps to account for burn-in. There were two sources in which the peak of the radio spectrum was completely unconstrained by the fitting procedure in both epochs, and one source where the peak of the spectrum was completely unconstrained for one epoch. For these fits we report a lower limit on the peak radio flux density and an upper limit on the synchrotron self-absorption break frequency. 

\begin{figure*}
    \centering
    \includegraphics[width=0.3\linewidth]{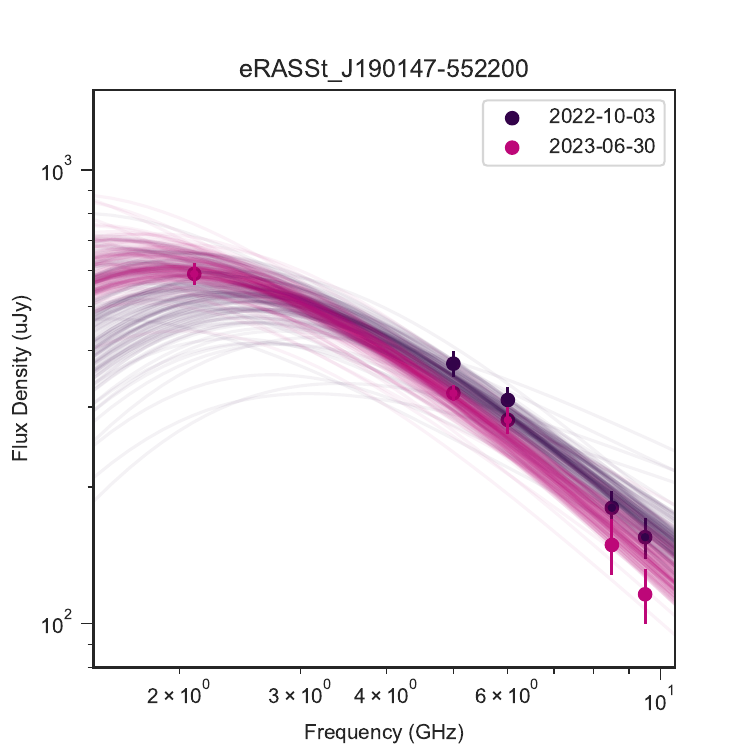}
        \includegraphics[width=0.3\linewidth]{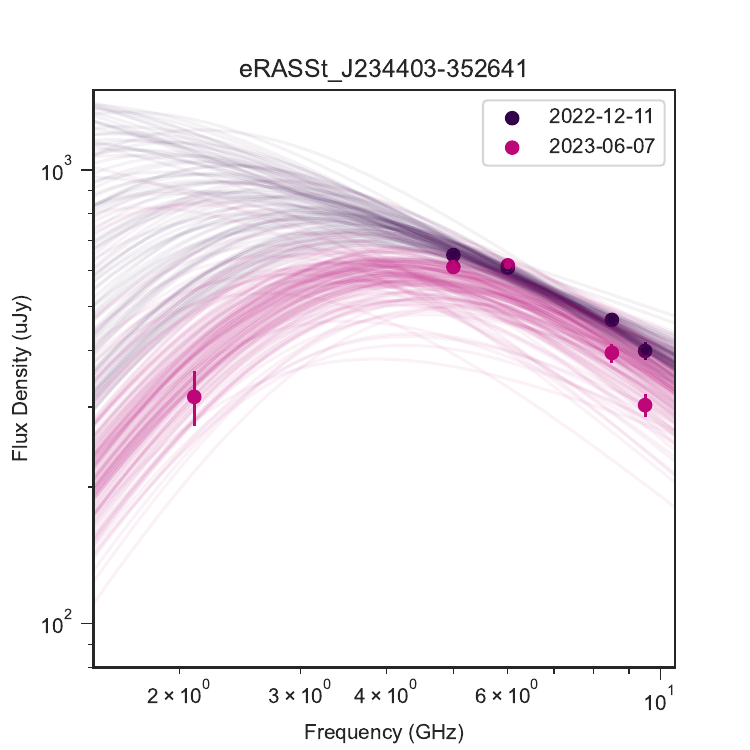}
        \includegraphics[width=0.3\linewidth]{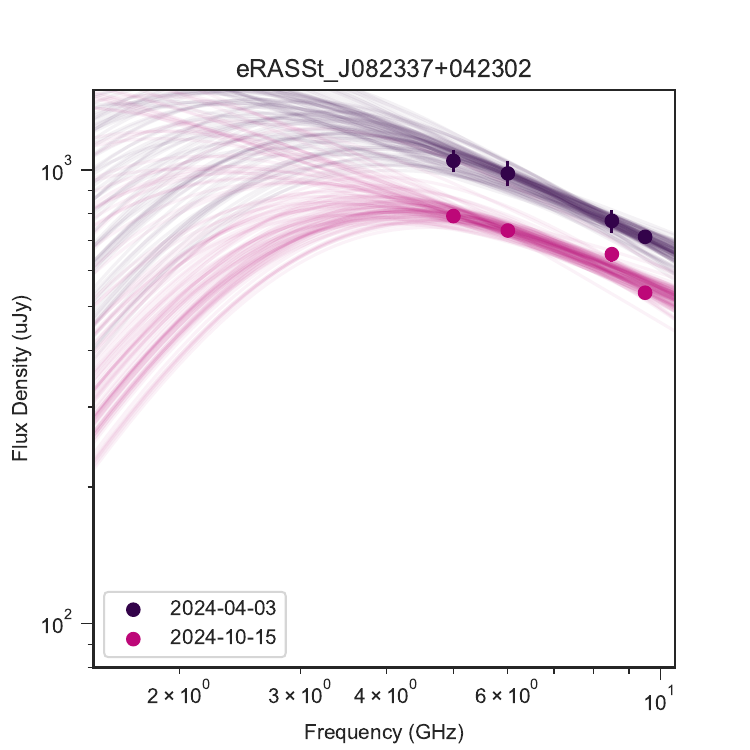}
        \includegraphics[width=0.3\linewidth]{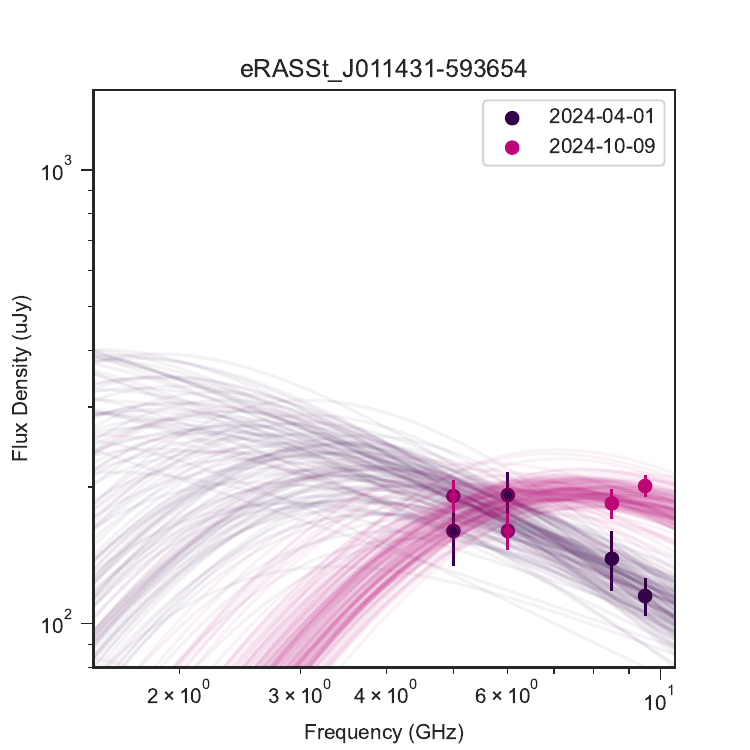}
        \includegraphics[width=0.3\linewidth]{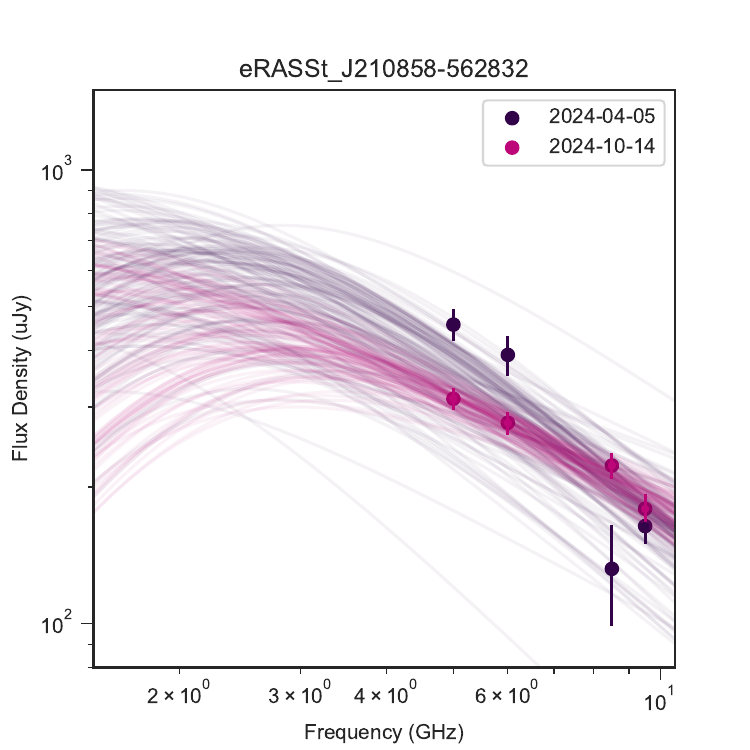}
        \includegraphics[width=0.3\linewidth]{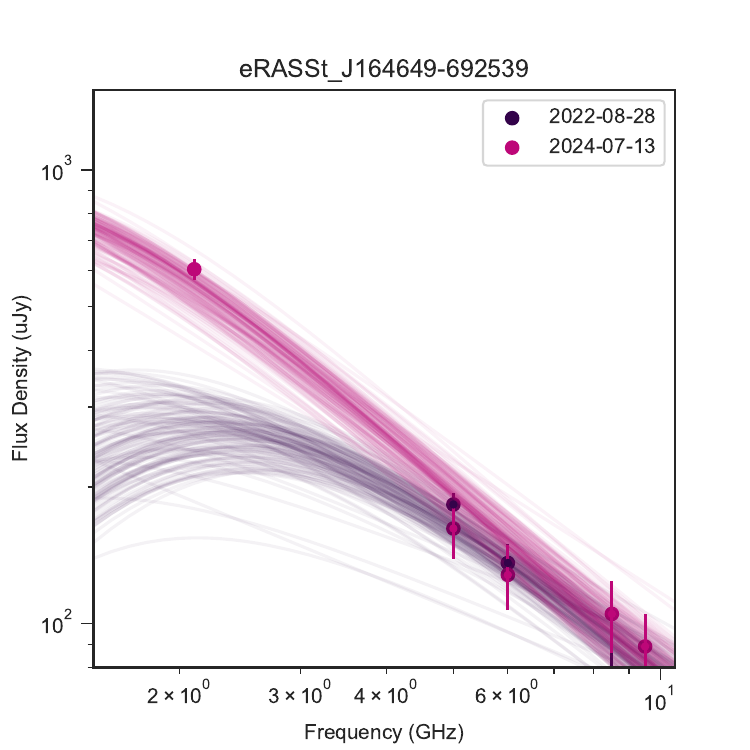}
    \caption{Synchrotron spectral fits for 6 of the TDEs in the sample with sufficient signal to extract radio spectra at each of the two epochs. The synchrotron model assumes the peak of the radio spectrum is associated with a self-absorption break, as appropriate for a TDE outflow. The spectral model and fitting method are described in Section \ref{sec:specfits}.}
    \label{fig:specfits}
\end{figure*}

The synchrotron spectral fit parameters are given in Table \ref{tab:specfits} and the resulting best-fit spectra are plotted in Figure \ref{fig:specfits}. Overall, we find best-fitting synchrotron indices of $p$ in the range 2.5--3.5 and peak frequencies in the range 1--5\,GHz, although note the peak frequencies are poorly constrained for most sources due to a lack of spectral coverage below 4.5\,GHz. These synchrotron fit parameters are typical for radio-emitting outflows from TDEs presented in the literature \citep[e.g.][]{Alexander2016,Goodwin2022,Cendes2021}.

\begin{sidewaystable*}[]
    \centering
        \caption{Radio spectral fit parameters and inferred outflow physical parameters for the TDE candidates with good radio spectral coverage.}
    \begin{tabular}{cccccccccccc}
\hline
Source & Epoch Date & $\Delta t$$^*$ (d) & $F_{\rm{ext}}$ (mJy) & $\nu_a$ (GHz) & $p$ &log $R_{\rm{eq}}$ (cm) & log $E_{\rm{eq}}$ (erg) & log $N_e$ & log $B$ (G) & log $M_{\rm{ej}}$ (g) & $v$ (c)\\
\hline
eRASSt J011431-593654 & 2024-04-01 & 1224 & $>$0.44 & $<$3.08 &     3.01$\pm$0.62 & $>$17.08 & $>$49.97    & 2.65$\pm$1.45 & -0.47$\pm$1.07 & 32.07$\pm$0.26 & 0.042$\pm$0.016\\
 & 2024-10-09 & 1415 & $>$0.44 & $<$5.14 &     2.86$\pm$0.60 & $>$16.85 & $>$49.61    & 3.00$\pm$0.71 & -0.30$\pm$0.52 & 32.29$\pm$0.16 & 0.022$\pm$0.004\\
\hline
eRASSt J082337+042302 & 2024-04-03 & 1251 & $>$2.57 & $<$2.60 &     2.79$\pm$0.53 & $>$16.83 & $>$48.99    & 2.42$\pm$1.32 & -0.59$\pm$1.03 & 31.69$\pm$0.24 & 0.021$\pm$0.008\\
 & 2024-10-15 & 1446 & $>$1.89 & $<$2.92 &     2.86$\pm$0.57 & $>$16.72 & $>$48.84    & 2.59$\pm$1.17 & -0.50$\pm$0.90 & 31.88$\pm$0.21 & 0.014$\pm$0.005\\
\hline
eRASSt J164649-692539 & 2022-08-28 & 702 & $>0.46$ & $<2.28$ &     3.23$\pm$0.50 & $>16.36$ & $>47.96$    & 2.81$\pm$1.41 & -0.39$\pm$0.99 & 31.11$\pm$0.24 & 0.013$\pm$0.005\\
 & 2024-07-13 & 1387 & $>1.54$ & $<0.99$ &     3.50$\pm$0.36 & $>16.99$ & $>49.17$    & 2.13$\pm$1.19 & -0.73$\pm$0.78 & 31.66$\pm$0.20 & 0.027$\pm$0.008\\
\hline
eRASSt J190147-552200 & 2022-10-03 & 908 & $>$0.99 & $<2.32$ &     3.29$\pm$0.43 & $>17.01$ & $>49.72$\    & 2.60$\pm$1.13 & -0.50$\pm$0.78 & 31.78$\pm$0.19 & 0.044$\pm$0.013\\
 & 2023-06-30 & 1178 & 1.64$\pm$0.13 & 1.45$\pm$0.22 &     3.52$\pm$0.26 & 17.34$\pm$0.07 & 50.37$\pm$0.08    & 2.28$\pm$0.64 & -0.66$\pm$0.42 & 32.03$\pm$0.10 & 0.070$\pm$0.011\\
\hline
eRASSt J210858-562832 & 2024-04-05 & 1257 & $>1.00$ & $<2.39$ & 3.21$\pm$0.60 & $>16.88$ & $>49.34$    & 2.63$\pm$2.03 & -0.48$\pm$1.43 & 31.94$\pm$0.35 & 0.024$\pm$0.012\\
 & 2024-10-14 & 1449 & $>0.81$ & $<2.46$ &   2.97$\pm$0.48 & $>16.81$ & $>49.02$    & 2.53$\pm$1.22 & -0.53$\pm$0.91 & 31.88$\pm$0.21 & 0.017$\pm$0.006\\
\hline
eRASSt J234403-352641 & 2022-12-11 & 746 & 1.73$\pm$0.17 & 2.60$\pm$0.67 &     3.05$\pm$0.41 & 17.27$\pm$0.11 & 50.34$\pm$0.12    & 2.44$\pm$0.99 & -0.58$\pm$0.73 & 31.72$\pm$0.17 & 0.096$\pm$0.025\\
 & 2023-06-07 & 924 & 1.51$\pm$0.29 & 3.13$\pm$0.46 &     3.34$\pm$0.48 & 17.19$\pm$0.08 & 50.43$\pm$0.12    & 2.79$\pm$0.71 & -0.40$\pm$0.47 & 32.14$\pm$0.14 & 0.066$\pm$0.011\\
\hline
    \end{tabular}
    \label{tab:specfits}
    \footnotesize{$^*$ $\Delta t$ is measured assuming $t_0$ is the time of the peak eROSITA X-ray flux measurement.}
    \footnotesize{$F_{ext}$ is the inferred synchrotron spectrum normalization, $\nu_a$ is the synchrotron self-absorption break frequency, $p$ is the synchrotron energy index, $R_{rm{eq}}$ is the inferred equipartition radius, $E_{\rm{eq}}$ is the inferred equipartition energy, $N_e$ the ambient electron density, $B$ is the magnetic field, $M_{\rm{ej}}$ is the mass in the radio-emitting region, and $v$ is the inferred outflow velocity.}
    
\end{sidewaystable*}

\subsubsection{Radio outflow properties}
For the 6 sources in which a synchrotron self-absorption break was able to be fit or an upper limit obtained in the radio spectra, and under the assumption the variable radio emission was produced by an outflow launched near the time of peak X-ray flux, we  can infer physical properties of the outflow. 

As in \citet{Goodwin2022,Goodwin2023,Goodwin2024}, we model a blast-wave in which the electrons are accelerated into a powerlaw distribution $N(\gamma)\propto \gamma^{-p}$. By assuming equipartition between the magnetic field and electron energy densities, we can derive physical outflow properties such as a minimum equipartition radius ($R_{eq}$), energy ($E_{eq}$), velocity ($v$), magnetic field strength ($B$), mass in the emitting region ($m_{ej}$), and ambient electron density ($n_e$). We follow the equipartition derivations outlined in \citet{BarniolDuran2013} and assume a spherical geometry of the outflow, to obtain the minimum radius and energy of the system. The exact equations we use to calculate $R_{eq}$, $E_{eq}$, $v$, $B$, $m_{ej}$, and $n_e$ are Equations 4--13 in \citet{Goodwin2022}. 

The inferred physical outflow properties for the 6 TDEs with detailed spectral modelling are listed in Table \ref{tab:specfits}. In general we find the minimum radii are in the range $10^{16.5}$--$10^{17.5}$\,cm with minimum energies $10^{49}$--$10^{50.5}$\,erg. We caution that the outflow parameters given represent the \textit{minimum} radius and energy for each source, and may deviate significantly from these values for different assumed outflow geometries as well as any deviation from equipartition. Due to the large uncertainties in the spectral fits we do not include additional models and instead report the equipartition radii and energies for a spherical geometry, which correspond to the minimum values of each. 

\section{Discussion}\label{sec:discussion}
Our results reveal 11 out of 22 X-ray selected TDE candidates in our sample were coincident with a radio source. 8 of these radio sources showed statistically significant variability over a 6 month period, and 6 showed radio variability greater than expected due to ISS of a compact source. Interestingly, all sources that showed statistically significant radio variability faded between the first and second epochs of observations. 

\subsection{Multiwavelength flares}
\textcolor{black}{\citet{Grotova_inpress}} searched available mid infra-red (IR) and optical archives for the presence of multiwavelength flares correlated with the X-ray flares observed. They found 5 of the TDE candidates in our sample had significant IR flares of more than 1 mag: J0439, J0644, J0823, J2108, and J2344. Additionally, 5 sources in our sample showed prominent optical flares: J0318, J0439, J0823, J1102, J2344, and J1421. 

Interestingly, 4 out of 5 of the sources with IR flares were radio-detected (although only 2 were variable), and 5 out of 6 of the sources with optical flares were radio-detected (although only 3 were variable), tentatively indicating that sources with a multiwavelength flare were more likely to be radio-detected.

\subsection{The nature of the radio emission from each source}
Radio emission observed from galaxies may be produced by a number of physical processes, including AGN activity, active star-forming regions, supernova remnants, or transient synchrotron-emitting events. The radio spectral properties, luminosity, and presence of variability allow for distinguishing between these physical processes \citep[e.g.][]{Ho2001,Hovatta2008}. Below we discuss the likely nature of each radio source in our sample based on the radio variability, luminosity, and spectral properties as well as the presence of multiwavelength flares. 

\subsubsection{Non-variable radio sources}
\begin{itemize}
    \item \textbf{eRASSt J110240-051813} was detected at both 5.5 and 9\,GHz but did not show statistically significant radio variability between the two observing epochs. The spectral index of the radio emission from this source of $\alpha=-0.7$ is entirely consistent with that expected for radio emission from star formation, new AGN activity, or synchrotron transient emission (as expected for a TDE). We note that the radio emission did fade over the 6 months between the two ATCA observing epochs, but the fading was not statistically significant due to the faintness of the source and the sensitivity of the observations. The X-ray flare was also quasi-simultaneous with an optical flare consistent with a TDE-like optical flare \textcolor{black}{(Grotova et al., submitted)}. We therefore cannot rule out that the radio emission is from a synchrotron-emitting TDE outflow, but require further observations. 
    \item \textbf{eRASSt J043959.5-651403} (also Gaia20cdq and AT2020jgh) and \textbf{eRASSt J063413-713908} were both detected at very faint radio flux densities ($<100$\,$\rm{\mu}$Jy, 4--5$\sigma$ detections). eRASSt J043959.5-651403 also showed an optical and infra-red flare \textcolor{black}{(Grotova et al.\ submitted)}. Without spectral information and further observations with a longer time baseline, it is difficult to conclude whether the radio emission from these sources is consistent with transient synchrotron emission or host galaxy emission. 
\end{itemize}

\subsubsection{Variable radio sources}
\begin{itemize}
    \item \textbf{eRASSt J210858-562832} showed variability at 9\,GHz greater than expected for ISS, implying there is likely some intrinsic radio variability. However, the radio spectrum is steep in the first epoch ($\alpha=-1.14$), and the synchrotron self-absorption model is a very poor fit to the data (Figure \ref{fig:specfits}), implying $p>3.5$, or the radio spectrum is dominated by a different physical process than synchrotron self-absorption. Interestingly, in the second epoch, the radio spectrum is flatter, and shows tentative evidence of a synchrotron self-absorption break (Figure \ref{fig:specfits}). The radio source is detected in ASKAP-RACS observations at 0.9 and 1.4\,GHz (Table \ref{tab:RACS_data}), but does not show statistically significant variability at 0.9\,GHz between 2019 and 2024. The first RACS observation is before the first eROSITA X-ray detection, which may indicate the radio source was present before the X-ray flare. The X-ray flare from this source was also quasi-simultaneous with an infra red flare \textcolor{black}{(Grotova et al.\ submitted)}. We deduce that the radio emission from this source is not consistent with ``normal" TDE radio-emitting outflow emission due to the steep initial spectrum and lack of low-frequency variability, but cannot provide strong conclusions about the nature of this radio source. 
    \item \textbf{eRASSt J190147-552200} also showed an optical flare quasi-simultaneous with the X-ray flare \textcolor{black}{(Grotova et al.\ submitted)}. The radio source shows a relatively steep spectrum ($\alpha=-1.5$ to $-1.9$) and the variability detected at 5.5 and 9\,GHz is consistent with ISS of a compact radio source. Although, the consistent temporal decay at both frequencies observed between the ATCA epochs (faded by 15--30$\%$) make ISS an unlikely explanation for the variability observed as the variability induced by ISS is stochastic. The synchrotron self-absorption model gives a reasonable fit to the data (Figure \ref{fig:specfits}), and the radio emission is consistent with a relatively large (old) TDE outflow, but still compact enough for ISS to affect the flux density over time. 
    \item \textbf{eRASSt J234403-352641} (also Gaia20eub and AT2020wjw) is a very strong TDE candidate presented by \citet{Homan2023} and the radio-emitting outflow was studied extensively by \citet{Goodwin2024}. The X-ray and radio flares were quasi-simultaneous with optical and infra-red flares making the transient very TDE-like, although an AGN ignition event cannot be ruled out \citep{Homan2023,Goodwin2024}. The radio emission from this source is well-fit by the synchrotron self-absorption model (Figure \ref{fig:specfits}), and the radio outflow properties are consistent with the broader TDE population \citep{Alexander2020}. 
    \item \textbf{eRASSt J082337+042302} is also known as AT2019avd, and was initially classified as a narrow-line Seyfert-1 galaxy due to strong Balmer lines, HeI emission lines, and an FeII complex \citep{Gezari2020}. Soft X-ray emission was discovered by eROSITA and the source showed Bowen fluorescence lines in the optical spectrum as well as an infra-red flare \citep{Malyali2021}. In the optical, the source is atypical of a TDE, showing two optical flares spaced approximately 200\,d apart \citep{Wang2023}. \citet{Malyali2021} propose the event could be a TDE or a more exotic type of nuclear transient. A new radio outflow was discovered and studied extensively by \citet{Wang2023}, in which they conclude the transient radio emission from this source is consistent with a compact radio-emitting outflow launched by a jet or disk wind from an atypical TDE. 
    The radio emission from this source is extremely variable, with the source undetected in 3\,GHz VLASS observations in 2017 and brightening drastically in VLASS observations between 2020 and 2023 (Table \ref{tab:VLASS_data}). The radio emission faded between epochs in our ATCA observations 1200-155\,d after the X-ray flare. We find the radio spectral index is consistent with synchrotron emission with a synchrotron spectral index $p\approx2.8$, and the lack of a peak in the spectrum $>4.5$\,GHz indicates a source size $>10^{16.8}$\,cm. The physical properties of this outflow are consistent with other TDE outflows observed to date. 
    \item \textbf{eRASSt J011431-593654} did not have flares detected at optical or IR wavelengths, and shows curious radio evolution. The radio variability is greater than expected for ISS, implying intrinsic variability from the radio source. The spectrum of the first epoch is consistent with synchrotron emission or host galaxy emission ($\alpha=-0.7$), however the spectrum became extremely flat 6 months later. The first epoch shows tentative indication of a peak between 5 and 6\,GHz, however this peak is not statistically significant given the errors on the flux density measurements. The flattening of the spectrum in the second epoch may be due to a compact jet in the system \citep{Blandford1979}. Blazars, which are widely accepted to be AGN with jets oriented towards the viewer, commonly show flat radio spectra \citep[e.g.][]{Abdo2010} and X-ray binaries in the low/hard state are typically observed to have flat radio spectra associated with compact jets \citep[e.g.][]{Fender2001}. For eRASSt J0114, the compact jet may have been launched by a TDE with a long-lived accretion flow, or from a pre-existing AGN in the galaxy. Only further multiwavelength observations of the source which constrain the galaxy type and radio spectral evolution will distinguish between these scenarios. 
    \item \textbf{eRASSt J142140-295321} showed a single optical transient detection quasi-simultaneous with the X-ray flare, that is possibly off-nuclear \textcolor{black}{(Grotova et al.\ submitted)}. The source is a candidate TDE around an IMBH (see discussion in the Appendix of \textcolor{black}{\citet{Grotova_inpress}}. The radio emission from this source is consistent with synchrotron self-absorption from a TDE outflow, and it faded beyond detection thresholds in the second epoch. 
    \item \textbf{eRASSt J143624-174105} was only detected at 5.5\,GHz in the first epoch, and faded beyond detection thresholds in the second epoch. Without radio spectral information it is difficult to conclude the nature of this radio source, other than the statistical significance of the variability observed. 
    \item \textbf{eRASSt J164649-692539} was detected in ASKAP-RACS observations at 0.9 and 1.4\,GHz. The radio source did not show statistically significant radio variability at 1.4\,GHz between Dec 2020 and Jan 2021. We detected the radio source in all ATCA observations at 2.1, 5.5, and 9\,GHz and detected small amplitude variability at 5.5 and 9\,GHz between the two epochs. The amplitude of radio variability detected is consistent with variability due to ISS of a compact radio source. Due to low amplitude of variability of the source we deduce the radio emission is likely host galaxy emission. We note the radio spectrum of this source is well-fit by a synchrotron self-absorption model, but the radio peak is below the observed band and is not well-constrained, so a single power-law is equally well-fit. 
\end{itemize}

\subsection{Comparison to other TDEs}
\begin{figure}
    \centering
    \includegraphics[width=\linewidth]{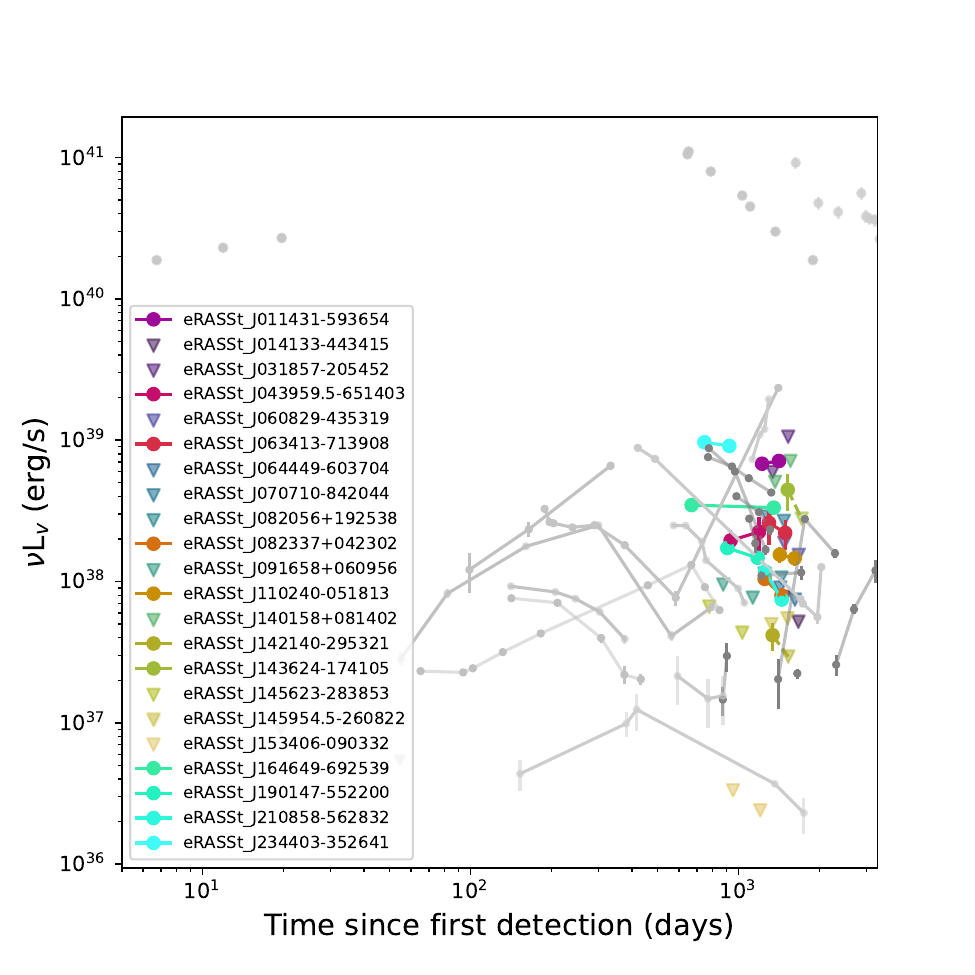}
    \caption{The radio luminosity with time since first detection of TDEs in the literature (grey points) and the 22 X-ray-selected TDE candidates in our sample (colored points). 3$\sigma$ upper limits are indicated by inverted triangles. The luminosity range of the X-ray-selected TDE candidates is consistent with the non-relativistic TDE population. The literature TDE data are from \citet{Cendes2024}, \citet{Goodwin2022}, \citet{Goodwin2023}, \citet{Alexander2016}, \citet{Cendes2021}, \citet{Horesh2021}, \citet{Horesh2021b}, \citet{Anderson2020}, \citet{Goodwin2023b}, \citet{Irwin2015}, and \citet{Zauderer2011}.}
    \label{fig:TDE_lc_comp}
\end{figure}

In Figure \ref{fig:TDE_lc_comp} we plot the radio lightcurves of known TDEs as well as our sample. In this figure it is clear that the radio emission luminosities and timescales of evolution from the eROSITA TDEs are broadly consistent with the non-relativistic radio-emitting population of events \citep[e.g.][]{Alexander2020,Cendes2021,Goodwin2022,Goodwin2023}. 
The lightcurves of the eROSITA TDEs occupy the region of time $1000-1500$\,d post discovery, which has been less well probed by observations currently available in the literature as it has only recently been discovered that TDEs are so long-lived in the radio. Our results confirm the findings of \citet{Cendes2024}: that transient radio emission from TDEs is extraordinarily long-lived. 

\subsubsection{Is there a difference between optical and X-ray selected TDEs in radio?}\label{sec:xraycomp}
Comparing the radio properties of an X-ray selected sample of TDEs with those of an optically selected sample of TDEs may provide insight into both the radio outflow ejection mechanism, as well as any possible difference between optically-bright and X-ray bright events.

In order to compare the radio properties of the X-ray selected sample with an optical sample, we choose the optically-selected sample from \citet{Cendes2024} to compare to as it also presents the radio properties of a sample of TDEs at late times ($500-3200$\,d after discovery). We find a remarkably similar detection rate of variable radio emission at times $>500$\,d post discovery for the optically and X-ray selected samples (40$\%$ in \citet{Cendes2024} vs 36$\%$ in our sample). Noting that although we detected statistically significant radio variability in 8 of our sources, we found the radio variability could not be ruled out as extrinsic due to ISS for 2 of those sources. 

\begin{figure}
    \centering
    \includegraphics[width=\linewidth]{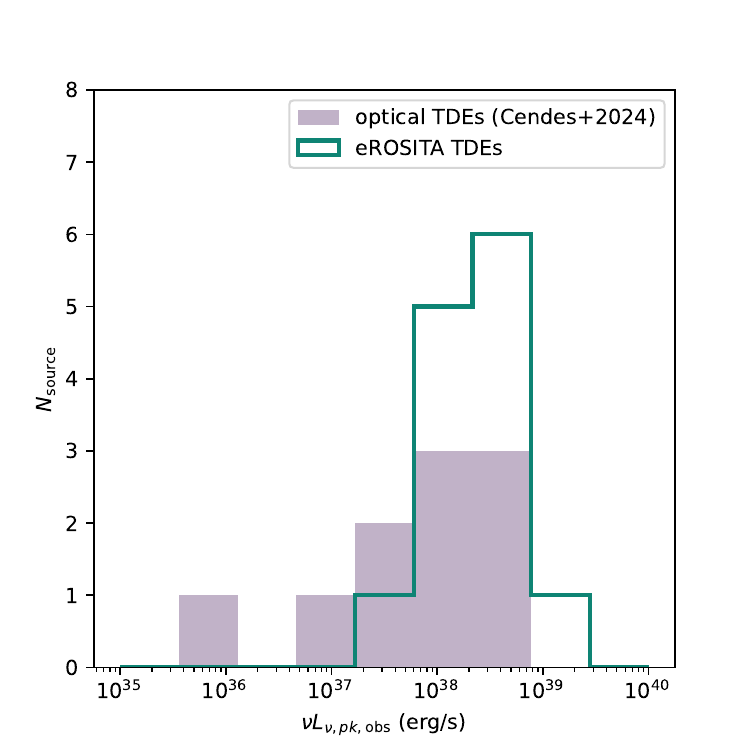}
    \caption{The observed peak 5.5\,GHz radio luminosity for the X-ray selected sample (eROSITA TDEs, green) compared to the optically-selected sample (optical TDEs, purple) from \citet{Cendes2024}. Note that the peak observed luminosities may not represent the true peak luminosities for each event. We find no statistically significant difference between the peak observed radio luminosity distribution of the two samples.}
    \label{fig:radL_hists}
\end{figure}

In Figure \ref{fig:radL_hists} we plot a histogram of the peak observed radio luminosity of each of the 11 radio-detected X-ray selected events presented in this work, compared to the peak radio luminosity of each of the detected optically-selected events from \citet{Cendes2024}. For both samples we have plotted all sources, regardless of whether the radio emission is variable or not. It is important to note that the peak \textit{observed} X-ray and radio luminosities may not correlate with the \textit{actual} peak X-ray and radio luminosities for these samples due to the paucity of the lightcurve data. To determine if there is a statistically significant difference between the radio luminosities of the optical and X-ray selected TDE samples, we calculate the KS-test statistic and p-value using the Python package \texttt{scipy.stats.ks2samp} for the two distributions. We find a KS-test statistic of 0.41 and p-value of 0.29, indicating that we cannot reject the null-hypothesis\footnote{In this work we assume a p-value $<0.05$ implies a statistically significant result.} that the two radio luminosity distributions are drawn from the same distribution. We therefore find no statistically significant difference between the radio luminosities and radio detection rate of the optically and X-ray selected samples.

\begin{figure*}
    \centering
    \includegraphics[width=0.45\linewidth]{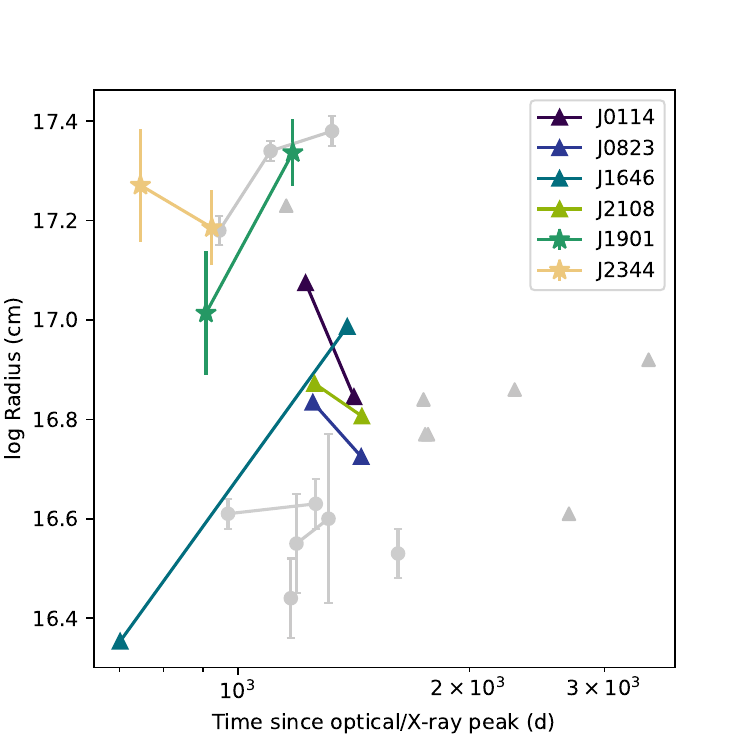}
    \includegraphics[width=0.45\linewidth]{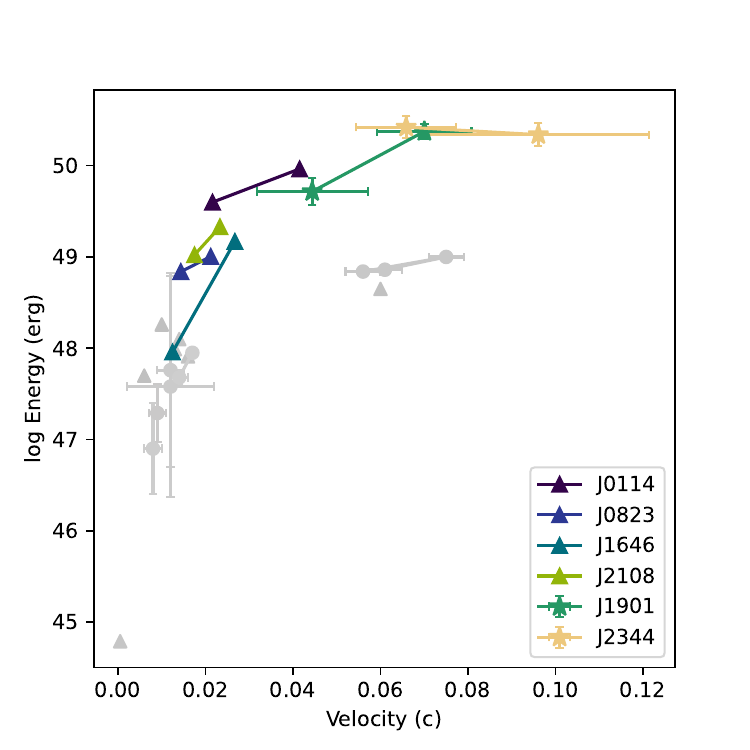}
    \caption{\textit{Left:} The equipartition radius plotted against time since peak optical or X-ray luminosity for the X-ray selected sample (colored stars) and the optically-selected sample from \citet{Cendes2024} (grey points). We find very similar ranges in equipartition radii for the two samples. \textit{Right:} The equipartition energy plotted against inferred outflow velocity for the X-ray selected sample (colored stars) and the optically-selected sample from \citet{Cendes2024} (grey points). We find a marginally statistically significant difference between the equipartition energies of the two samples, with X-ray selected events tending to have higher outflow equipartition energy for a similar distribution of velocities.}
    \label{fig:tde_outflow_comp}
\end{figure*}

Next, we compare the inferred physical outflow properties of the 6 X-ray selected TDEs we were able to carry out spectral modelling of to the physical outflow properties of the optically-selected TDEs from \citet{Cendes2024}. In Figure \ref{fig:tde_outflow_comp} (left panel) we compare the inferred equipartition radii over time for the two samples and in Figure \ref{fig:tde_outflow_comp} (right panel) we compare the inferred equipartition energies and velocities. Due to the lack of spectral coverage below 4.5\,GHz for the X-ray selected sample, the equipartition radii and energies reported represent the absolute minimum values, and may significantly change from the plotted values depending on the frequency of the self-absorption break. Nevertheless, we find the inferred equipartition radii are very similar for both the optical and X-ray samples. We find the X-ray selected sample tends to have higher minimum equipartition energies for a similar range in outflow velocities when compared to the optically-selected sample. Comparing the X-ray and optically-selected sample energy distributions, we calculate a KS-test statistic of 0.74 and p-value 0.01, rejecting the null-hypothesis that they are drawn from the same energy distribution. However, we note that this result is borderline significant, and a larger sample of sources is required to confirm the statistical significance of this trend. 

\subsubsection{Electron power-law index distribution}
\begin{figure}
    \centering
    \includegraphics[width=\linewidth]{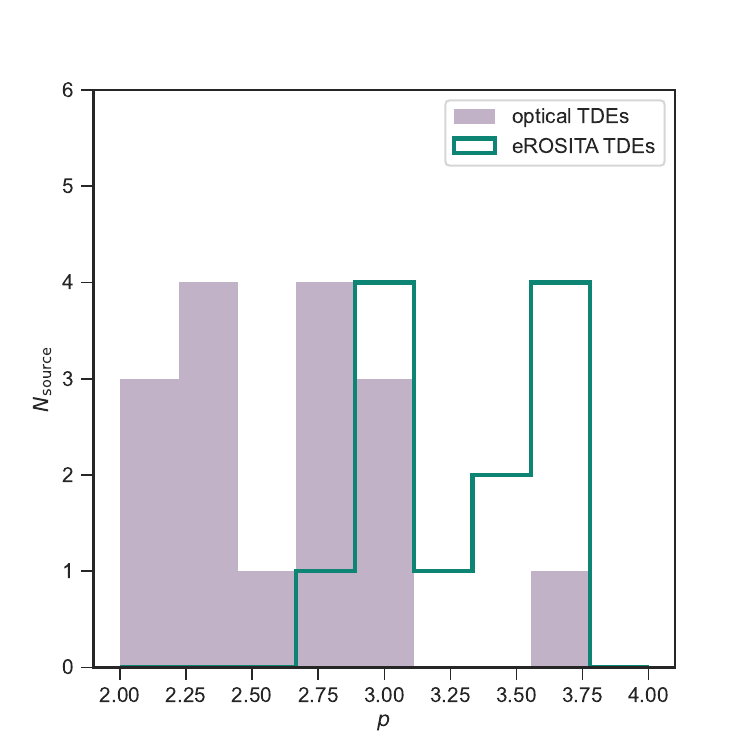}
    \caption{The fitted synchrotron spectral index, $p$, for the spectra of the X-ray selected TDEs (green) and the optically-selected sample from \citet{Cendes2024} (purple). A KS-test indicates a statistically significant difference between the spectral indices of the two samples despite both being fitted allowing the same range of $p$ ($2 < p < 4$), with X-ray selected events tending to have higher values of $p$.}
    \label{fig:pdist}
\end{figure}

Next we compare the modelled values of the synchrotron electron energy index, $p$, for each spectrum modelled (noting that the same source may appear multiple times if multiple spectra were available for that source). In Figure \ref{fig:pdist} we plot $p$ for each of the X-ray selected TDE spectra compared to each of the optically selected TDE spectra fitted in \citet{Cendes2024}. As above, we perform a KS-test to determine if the two samples are consistent with coming from the same distribution. We calculate a KS-test statistic of 0.79 and p-value 0.0001, rejecting the null-hypothesis with confidence that they are drawn from the same energy distribution.

Overall, the electron energy index for the X-ray selected sample has mean $3.3\pm0.3$, whilst the optically selected sample has mean $2.6\pm0.4$. Recall that in a synchrotron-emitting shock, the electrons are accelerated into a powerlaw distribution, $N(\gamma)\propto\gamma^{-p}$ \citep[e.g.][]{Granot2002}. We therefore deduce that the X-ray selected events may produce synchrotron emission with a steeper power-law index. A steeper power-law index indicates greater synchrotron losses, which may indicate the synchrotron-emitting region is older in these sources \citep[e.g.][]{Krolik1991}. This is consistent with the observation that all 8 variable radio sources in the X-ray sample were fading at times $>1000$\,d post-X-ray discovery, compared to 5 out of 11 of the optically-selected sample. 

\subsection{Radio to X-ray correlation}
\begin{figure}
    \centering
    \includegraphics[width=\linewidth]{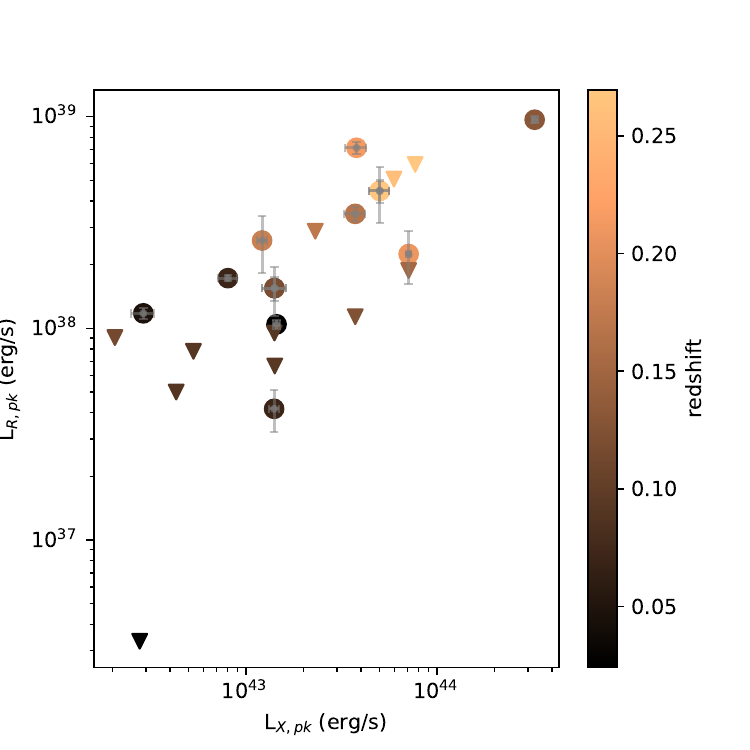}
    \caption{The peak observed 5.5\,GHz radio luminosity plotted against the peak observed X-ray luminosity for each of the 22 TDE candidates in the sample. Note that the peak observed luminosities may not represent the true peak luminosities for each event. Each point is coloured by the redshift of the host galaxy. Inverted triangles indicate 3$\sigma$ radio luminosity upper limits. A correlation between the peak radio and X-ray luminosities is apparent, however we caution that there is also a strong distance dependence of this correlation.}
    \label{fig:lx_v_lr}
\end{figure}

We searched for any correlation between the radio and X-ray luminosities of the TDEs in our sample. In Figure \ref{fig:lx_v_lr} we plot the peak observed radio 5.5\,GHz luminosity against the peak observed X-ray luminosity for each of the TDEs in the sample. Again, it is important to note that the peak \textit{observed} X-ray and radio luminosities may not correlate with the \textit{actual} peak X-ray and radio luminosities for these samples due to the sparse sampling of the lightcurve data. In Figure \ref{fig:lx_v_lr} a positive trend is apparent, where more luminous X-ray emission is associated with more luminous radio emission. However, we note that the trend may be driven by distance, as the most luminous events are at the largest distances. A larger, complete, sample of nearby events would help confirm if this positive correlation between radio and X-ray luminosities is real. 

\subsection{Implications for the outflow mechanism}
In the scenario in which the positive correlation we find between X-ray and radio luminosity is intrinsic to TDEs, this may imply that the radio outflow ejection mechanism is linked to the physical process that produces the X-ray emission in TDEs. 
For example, as mentioned in Section \ref{sec:intro}, for X-ray binaries and AGN, a positive observational correlation between X-ray and radio luminosity is well established during the hard state when a compact radio jet is present \citep[e.g.][]{Gallo2003,Brinkmann2000}. 
However, this correlation has not been seen for the transient radio-emitting ejecta that are released during the low state of X-ray binaries. 

For TDEs the mechanism that powers the X-ray emission is less well-established. Most X-ray detected TDEs show super soft X-ray spectra, unlike X-ray emission usually associated with powerful jets. Very rarely ($\sim1\%$ of events), a clear observational signal for a relativistic jet is seen from X-ray through to radio wavelengths \citep[e.g.][]{Bloom2011,Burrows2011,Levan2011,Zauderer2011,Andreoni2022,Pasham2023}.
It is unlikely that all TDEs produce these powerful jets \citep[e.g.][]{vanvelzen2016}, implying the conditions for launching powerful jets are only reached rarely. \citet{Dai2018} proposed that the differences in jet properties/detection rates and presence of optical or X-ray emission may be explained by viewing angle differences between TDEs. However, if all events produced powerful relativistic jets, we would expect to detect a large number of off-axis relativistic jets which may manifest as a large number of late-time brightening events \citep[e.g.][]{Matsumoto2023}. In an alternate scenario, the radio and X-ray correlation may be driven by less powerful, sub-relativistic jets, as was originally proposed for the radio outflow from ASASSN-14li \citep{vanvelzen2016}. These sub-relativistic jets may be launched near peak accretion in the disk, when the mass accretion rate is sufficiently high to trigger a jet launching episode. Alternatively, in the cooling envelope model \citep{Metzger2022}, the X-ray and optical emission is powered by reprocessing in a large envelope structure formed shortly after the stellar disruption. \citet{Hu2024} showed in numerical simulations of TDEs that this envelope-like structure contains some unbound material, which may provide sufficient mass to produce radio synchrotron emission as it shocks the CNM. In this scenario, a correlation between radio and X-ray luminosities may also be produced. Therefore, the presence of a tentative correlation between X-ray and radio luminosities in the events studied in our sample provides little insight into the physical outflow mechanism, other than that there may be a link between the physical processes that drive the radio and X-ray emission. 

\subsection{On the lack of late-time rising radio emission in the X-ray selected sample}
Interestingly, of the 8 variable radio sources in the X-ray selected sample, we found none with radio emission rising at $>1000$\,d post-X-ray discovery. In contrast, 5 out of 11 of the radio variable TDEs in the optically selected sample were rising at times $>1000$\,d post-optical discovery \citep{Cendes2024}. \textcolor{black}{\citet{Grotova_inpress}} noticed that in the X-ray selected TDE sample, the peak of the X-ray lightcurve was often significantly delayed from the optical or infra-red lightcurve peak in events in which a multiwavelength flare was observed. As we measure time since X-ray peak for our sample, compared to the \citet{Cendes2024} optical sample in which time is measured since optical discovery, this may explain why there are no apparent late-time flares in our sample. Although this is not always the case, as for AT2022dsb \citet{Malyali2023b} discovered a bright X-ray flare that rapidly dimmed pre-optical peak. We propose that the significance of no late-time rising radio emission in our X-ray selected sample may imply that many radio-emitting outflows may be launched near the X-ray lightcurve peak, which can be significantly delayed from the optical/IR lightcurve peaks.
This suggestion has recently gained some traction with \citet{Hajela2024} finding evidence for the second radio flare from ASASSN-15oi being launched near peak accretion rate, and \citet{Goodwin2024b} suggesting the same for the AT2020vwl. Further observations that investigate any correlation between X-ray peak and radio launching times will constrain whether this tentative suggestion is true for most TDEs.

\section{Summary}\label{sec:summary}
In this study, we analysed the radio properties of a systematic X-ray selected sample of TDEs discovered by the eROSITA telescope. Our conclusions are summarised as follows:
\begin{itemize}
    \item Of the 22 TDE candidates in our sample, 11 were coincident with a radio source, 8 (36$\%$) showed statistically significantly variable radio emission over a period of 6 months, and 6 (27$\%$) showed radio variability greater than expected for interstellar scintillation.
    \item We found no statistically significant difference between the detection rate and radio luminosities of the X-ray selected sample when compared to an optically selected sample \citep{Cendes2024} at similar times post-discovery.
    \item We found similar physical outflow properties between the X-ray and optically selected samples, such as minimum equipartition radius and velocity, although note that the synchrotron self-absorption turnover is not resolved for most sources in the X-ray sample and thus the inferred outflow properties are significantly uncertain. We tentatively found the X-ray selected events tend to have more energetic outflows than the optically-selected events with marginal statistical significance (KS-test statistic of 0.74, see Section \ref{sec:xraycomp}).
    \item We calculate a mean synchrotron electron spectral index of $p=3.3\pm0.3$ for the sample presented in this work. On comparison with the electron spectral indices of the optically-selected sample of TDEs (which has mean synchrotron electron spectral index of $2.6\pm0.4$), we find the X-ray selected sample has statistically significantly higher values of $p$, which comparisons of larger sample sizes may confirm. 
    \item Interestingly, all of the 8 variable radio sources in the X-ray selected sample were fading at times $>1000$\,d post-X-ray-discovery, compared to the optically-selected sample in which 5 out of 11 radio variable sources were rising in the radio at times $>1000$\,d post-optical-discovery. We suggest this lack of late-time rising radio emission from X-ray selected TDEs may be due to outflows that are launched near peak accretion rate in the disk (i.e. peak X-ray luminosity) which can be significantly delayed from the peak optical or IR luminosities \citep{Grotova2025}, in contrast to outflows that may be launched by debris collisions during the debris circularisation. 
    \item Finally, we found a tentative positive trend between the peak observed radio and X-ray luminosities in our sample, which may indicate the same underlying physical process powers X-ray and radio emission in TDEs. However, there is a strong distance dependence in this correlation and a larger sample is required to confirm the statistical significance of this correlation.
\end{itemize}

Overall this study has confirmed findings from previous studies that radio emission from TDEs is common (approximately half of all events show detectable radio-emitting outflows), and is extraordinarily long-lived, lasting years after the initial flare. These findings motivate further long-term and late-time TDE radio follow-up campaigns to gain insight into the ejection of outflows from black hole accretion events. The Einstein-Probe X-ray mission \citep{Yuan2015} will be crucial in increasing the sample of nearby X-ray selected TDEs in the near future, whilst the SKA-mid and ngVLA will be crucial in characterising radio outflows from TDEs on a large scale. 

\begin{acknowledgments}
AG is grateful for support from the Forrest Research Foundation. This work was supported by the Australian government through the Australian Research Council’s Discovery Projects funding scheme (DP200102471). The Australia Telescope Compact Array is part of the Australia Telescope National Facility (grid.421683.a) which is funded by the Australian Government for operation as a National Facility managed by CSIRO. We acknowledge the Gomeroi people as the traditional owners of the Observatory site.
\end{acknowledgments}

\vspace{5mm}
\facilities{ATCA, eROSITA, ASKAP, VLA}

\software{astropy \citep{2013A&A...558A..33A,2018AJ....156..123A},  
          }

\appendix

\section{Appendix information}

\subsection{Archival radio flux density measurements}

\begin{longtable}{lllll}
\caption{ASKAP RACS low (0.9\,GHz) and mid (1.4\,GHz) observations of the 22 sources in our sample between 2019--2024.}\\
\hline
Source & Date (UTC) & Date (MJD) & Frequency (GHz) & Flux Density ($\rm{\mu}$Jy)\\
\hline
Undetected sources & & & & \\
\hline
eRASSt J011431-593654 & & & &\\
 & 2020-03-28 & 58936 & 0.89& $<$1390\\
  & 2021-01-19 & 59233 & 1.37& $<$523\\
\hline
eRASSt J014133-443415 & & & &\\
 & 2019-04-30 & 58603 & 0.89& $<$639\\
 & 2020-12-31 & 59214 & 1.37& $<$500\\
 & 2021-01-10 & 59224 & 1.37& $<$1688\\
 & 2024-01-09 & 60318 & 0.94& $<$526\\
\hline
eRASSt J031857-205452 & & & &\\
 & 2021-01-20 & 59234 & 1.37& $<$661\\
& 2024-01-04 & 60314 & 0.94& $<$658\\
\hline
eRASSt J043959.5-651403 & & & &\\
 & 2020-05-02 & 58971 & 0.89& $<$1034\\
  & 2021-02-05 & 59250 & 1.37& $<$527\\
\hline
eRASSt J060829-435319 & & & &\\
 & 2020-03-28 & 58936 & 0.89& $<$530\\
 & 2021-01-26 & 59241 & 1.37& $<$477\\
  & 2024-01-06 & 60316 & 0.94& $<$500\\
\hline
eRASSt J063413-713908 & & & &\\
 & 2019-10-31 & 58788 & 0.89& $<$514\\
 & 2021-02-06 & 59252 & 1.37& $<$490\\
 & 2024-01-12 & 60322 & 0.94& $<$534\\
\hline
eRASSt J064449-603704 & & & &\\
 & 2021-02-05 & 59251 & 1.37& $<$499\\
  & 2024-01-09 & 60319 & 0.94& $<$456\\
\hline
eRASSt J070710-842044 & & & &\\
 & 2021-02-06 & 59252 & 1.37& $<$492\\
 & 2024-01-13 & 60323 & 0.94& $<$583\\
\hline
eRASSt J082056+192538 & & & &\\
 & 2021-01-11 & 59226 & 1.37& $<$518\\
 & 2023-12-30 & 60309 & 0.94& $<$712\\
 & 2024-01-26 & 60336 & 0.94& $<$524\\
\hline
eRASSt J091658+060956 & & & &\\
 & 2020-06-21 & 59021 & 0.89& $<$511\\
 & 2021-01-17 & 59232 & 1.37& $<$515\\
\hline
eRASSt J110240-051813 & & & &\\
 & 2020-05-02 & 58972 & 0.89& $<$549\\
 & 2021-01-17 & 59232 & 1.37& $<$493\\
\hline
eRASSt J140158+081402 & & & &\\
 & 2019-04-24 & 58598 & 0.89& $<$777\\
 & 2021-01-02 & 59217 & 1.37& $<$502\\
 & 2024-01-14 & 60324 & 0.94& $<$630\\
\hline
eRASSt J142140-295321 & & & &\\
 & 2020-03-26 & 58935 & 0.89& $<$905\\
  & 2021-01-17 & 59232 & 1.37& $<$521\\
\hline
eRASSt J143624-174105 & & & &\\
 & 2021-02-24 & 59270 & 1.37& $<$683\\
 & 2021-07-29 & 59424 & 1.37& $<$854\\
 & 2022-05-30 & 59730 & 1.37& $<$655\\
 & 2024-01-26 & 60336 & 0.94& $<$563\\
\hline
eRASSt J143815-203909 & & & &\\
 & 2021-02-24 & 59270 & 1.37& $<$1602\\
  & 2022-05-25 & 59725 & 1.37& $<$1613\\
& 2021-07-29 & 59424 & 1.37& $<$572\\
  & 2024-01-26 & 60336 & 0.94& $<$848\\
\hline
eRASSt J145623-283853 & & & &\\
 & 2020-03-26 & 58935 & 0.89& $<$589\\
 & 2021-01-17 & 59232 & 1.37& $<$472\\
\hline
eRASSt J190147-552200 & & & &\\
 & 2019-05-06 & 58610 & 0.89& $<$847\\
 & 2020-12-26 & 59209 & 1.37& $<$464\\
 & 2023-12-30 & 60308 & 0.94& $<$463\\
\hline
eRASSt J234403-352641 & & & &\\
 & 2021-01-10 & 59224 & 1.37& $<$605\\
 & 2024-01-07 & 60316 & 0.94& $<$994\\
\hline
eRASSt J145954.5-260822 & & & &\\
& 2021-01-17 & 59232 & 1.37& $<486$\\
\hline
Detected sources & & & & \\
\hline
eRASSt J082337+042302 & & & &\\
 & 2021-01-17 & 59232 & 1.37& 862$\pm$221\\
 & 2023-12-31 & 60310 & 0.94& 1335$\pm$277\\
\hline
eRASSt J164649-692539 & & & &\\
 & 2020-03-26 & 58935 & 0.89& 1888$\pm$162\\
 & 2020-12-28 & 59211 & 1.37& 1221$\pm$160\\
 & 2021-01-17 & 59231 & 1.37& 2222$\pm$860\\
\hline
eRASSt J210858-562832 & & & &\\
& 2019-05-07 & 58610& 0.89& 1330$\pm$220\\
& 2020-03-28 & 58936 & 0.89& 989$\pm$251\\
& 2021-01-06 & 59220 & 1.37& 952$\pm$167\\
& 2024-01-21 & 60330 & 0.89& 1170$\pm$150\\
\hline
    \label{tab:RACS_data}
\end{longtable}

\begin{longtable}{llll}
\caption{VLASS 3\,GHz observations taken between 2017--2024 of 17 of the sources in our sample that are covered by the VLASS footprint.}\\
\hline
Source & Date (UTC) & Date (MJD) & Flux Density ($\rm{\mu}$Jy)\\
\hline
Undetected Sources & & & \\
\hline
eRASSt J031857-205452 & & &\\
 & 2019-06-29 & 58664 & $<$458\\
 & 2022-02-12 & 59622 & $<$377\\
\hline
eRASSt J082056+192538 & & &\\
 & 2019-04-14 & 58587 & $<$354\\
 & 2021-11-17 & 59535 & $<$330\\
 & 2024-05-02 & 60432 & $<$399\\
\hline
eRASSt J091658+060956 & & &\\
 & 2017-11-21 & 58078 & $<$383\\
  & 2020-08-08 & 59070 & $<$513\\
  & 2023-02-16 & 59991 & $<$452\\
\hline
eRASSt J110240-051813 & & &\\
 & 2019-04-02 & 58575 & $<$440\\
& 2024-06-16 & 60477 & $<$417\\
 & 2021-11-16 & 59535 & $<$423\\
\hline
eRASSt J140158+081402 & & &\\
 & 2019-05-08 & 58611 & $<$578\\
 & 2021-10-04 & 59492 & $<$432\\
  & 2024-06-27 & 60488 & $<$425\\
\hline
eRASSt J142140-295321 & & &\\
 & 2019-06-29 & 58663 & $<$482\\
 & 2022-02-08 & 59618 & $<$384\\
\hline
eRASSt J143624-174105 & & &\\
 & 2018-02-16 & 58165 & $<$400\\
 & 2020-09-18 & 59111 & $<$495\\
 & 2023-06-12 & 60107 & $<$463\\
\hline
eRASSt J143815-203909 & & &\\
 & 2018-02-12 & 58162 & $<$389\\
 & 2020-09-05 & 59147 & $<$425\\
 & 2023-06-19 & 60114 & $<$402\\
\hline
eRASSt J145623-283853 & & &\\
 & 2019-07-02 & 58666 & $<$475\\
 & 2022-02-06 & 59617 & $<$377\\
\hline
eRASSt J234403-352641 & & &\\
 & 2018-02-11 & 58161 & $<$352\\
 & 2020-11-05 & 59158 & $<$480\\
 & 2023-06-05 & 60101 & $<$392\\
\hline
eRASSt J145954.5-260822 & & &\\
 & 2019-07-12 & 58676 & $<$453\\
 & 2022-02-05 & 59616 & $<$382\\
\hline
Detected Sources & & & \\
\hline
eRASSt J082337+042302 & & &\\
 & 2017-11-18 & 58075 & $<$404\\
 & 2020-08-09 & 59071 & 607$\pm$170$^*$\\
 & 2023-01-24 & 59968 & 2800$\pm$250\\
\hline
    \label{tab:VLASS_data}
\footnotesize{$^*$Marginal detection ($<5\sigma$ but $>3\sigma$).}
\end{longtable}

\bibliography{bibfile}{}

\begin{thebibliography}{}
\expandafter\ifx\csname natexlab\endcsname\relax\def\natexlab#1{#1}\fi
\providecommand{\url}[1]{\href{#1}{#1}}
\providecommand{\dodoi}[1]{doi:~\href{http://doi.org/#1}{\nolinkurl{#1}}}
\providecommand{\doeprint}[1]{\href{http://ascl.net/#1}{\nolinkurl{http://ascl.net/#1}}}
\providecommand{\doarXiv}[1]{\href{https://arxiv.org/abs/#1}{\nolinkurl{https://arxiv.org/abs/#1}}}

\bibitem[{{Abdo} {et~al.}(2010){Abdo}, {Ackermann}, {Agudo}, {Ajello}, {Aller}, {Aller}, {Angelakis}, {Arkharov}, {Axelsson}, {Bach}, {Baldini}, {Ballet}, {Barbiellini}, {Bastieri}, {Baughman}, {Bechtol}, {Bellazzini}, {Benitez}, {Berdyugin}, {Berenji}, {Blandford}, {Bloom}, {Boettcher}, {Bonamente}, {Borgland}, {Bregeon}, {Brez}, {Brigida}, {Bruel}, {Burnett}, {Burrows}, {Buson}, {Caliandro}, {Calzoletti}, {Cameron}, {Capalbi}, {Caraveo}, {Carosati}, {Casandjian}, {Cavazzuti}, {Cecchi}, {{\c{C}}elik}, {Charles}, {Chaty}, {Chekhtman}, {Chen}, {Chiang}, {Chincarini}, {Ciprini}, {Claus}, {Cohen-Tanugi}, {Colafrancesco}, {Cominsky}, {Conrad}, {Costamante}, {Cutini}, {D'ammando}, {Deitrick}, {D'Elia}, {Dermer}, {de Angelis}, {de Palma}, {Digel}, {Donnarumma}, {Silva}, {Drell}, {Dubois}, {Dultzin}, {Dumora}, {Falcone}, {Farnier}, {Favuzzi}, {Fegan}, {Focke}, {Forn{\'e}}, {Fortin}, {Frailis}, {Fuhrmann}, {Fukazawa}, {Funk}, {Fusco}, {G{\'o}mez}, {Gargano}, {Gasparrini}, {Gehrels}, {Germani}, {Giebels}, {Giglietto},
  {Giommi}, {Giordano}, {Giuliani}, {Glanzman}, {Godfrey}, {Grenier}, {Gronwall}, {Grove}, {Guillemot}, {Guiriec}, {Gurwell}, {Hadasch}, {Hanabata}, {Harding}, {Hayashida}, {Hays}, {Healey}, {Heidt}, {Hiriart}, {Horan}, {Hoversten}, {Hughes}, {Itoh}, {Jackson}, {J{\'o}hannesson}, {Johnson}, {Johnson}, {Jorstad}, {Kadler}, {Kamae}, {Katagiri}, {Kataoka}, {Kawai}, {Kennea}, {Kerr}, {Kimeridze}, {Kn{\"o}dlseder}, {Kocian}, {Kopatskaya}, {Koptelova}, {Konstantinova}, {Kovalev}, {Kovalev}, {Kurtanidze}, {Kuss}, {Lande}, {Larionov}, {Latronico}, {Leto}, {Lindfors}, {Longo}, {Loparco}, {Lott}, {Lovellette}, {Lubrano}, {Madejski}, {Makeev}, {Marchegiani}, {Marscher}, {Marshall}, {Max-Moerbeck}, {Mazziotta}, {McConville}, {McEnery}, {Meurer}, {Michelson}, {Mitthumsiri}, {Mizuno}, {Moiseev}, {Monte}, {Monzani}, {Morselli}, {Moskalenko}, {Murgia}, {Nestoras}, {Nilsson}, {Nizhelsky}, {Nolan}, {Norris}, {Nuss}, {Ohsugi}, {Ojha}, {Omodei}, {Orlando}, {Ormes}, {Osborne}, {Ozaki}, {Pacciani}, {Padovani}, {Pagani}, {Page},
  {Paneque}, {Panetta}, {Parent}, {Pasanen}, {Pavlidou}, {Pelassa}, {Pepe}, {Perri}, {Pesce-Rollins}, {Piranomonte}, {Piron}, {Pittori}, {Porter}, {Puccetti}, {Rahoui}, {Rain{\`o}}, {Raiteri}, {Rando}, {Razzano}, {Reimer}, \& {Reimer}}]{Abdo2010}
{Abdo}, A.~A., {Ackermann}, M., {Agudo}, I., {et~al.} 2010, \apj, 716, 30, \dodoi{10.1088/0004-637X/716/1/30}

\bibitem[{{Alexander} {et~al.}(2016){Alexander}, {Berger}, {Guillochon}, {Zauderer}, \& {Williams}}]{Alexander2016}
{Alexander}, K.~D., {Berger}, E., {Guillochon}, J., {Zauderer}, B.~A., \& {Williams}, P.~K.~G. 2016, \apjl, 819, L25, \dodoi{10.3847/2041-8205/819/2/L25}

\bibitem[{{Alexander} {et~al.}(2020){Alexander}, {van Velzen}, {Horesh}, \& {Zauderer}}]{Alexander2020}
{Alexander}, K.~D., {van Velzen}, S., {Horesh}, A., \& {Zauderer}, B.~A. 2020, \ssr, 216, 81, \dodoi{10.1007/s11214-020-00702-w}

\bibitem[{{Anderson} {et~al.}(2020){Anderson}, {Mooley}, {Hallinan}, {Dong}, {Phinney}, {Horesh}, {Bourke}, {Cenko}, {Frail}, {Kulkarni}, \& {Myers}}]{Anderson2020}
{Anderson}, M.~M., {Mooley}, K.~P., {Hallinan}, G., {et~al.} 2020, \apj, 903, 116, \dodoi{10.3847/1538-4357/abb94b}

\bibitem[{{Andreoni} {et~al.}(2022){Andreoni}, {Coughlin}, {Perley}, {Yao}, {Lu}, {Cenko}, {Kumar}, {Anand}, {Ho}, {Kasliwal}, {de Ugarte Postigo}, {Sagu{\'e}s-Carracedo}, {Schulze}, {Kann}, {Kulkarni}, {Sollerman}, {Tanvir}, {Rest}, {Izzo}, {Somalwar}, {Kaplan}, {Ahumada}, {Anupama}, {Auchettl}, {Barway}, {Bellm}, {Bhalerao}, {Bloom}, {Bremer}, {Bulla}, {Burns}, {Campana}, {Chandra}, {Charalampopoulos}, {Cooke}, {D'Elia}, {Das}, {Dobie}, {Ag{\"u}{\'\i} Fern{\'a}ndez}, {Freeburn}, {Fremling}, {Gezari}, {Goode}, {Graham}, {Hammerstein}, {Karambelkar}, {Kilpatrick}, {Kool}, {Krips}, {Laher}, {Leloudas}, {Levan}, {Lundquist}, {Mahabal}, {Medford}, {Miller}, {M{\"o}ller}, {Mooley}, {Nayana}, {Nir}, {Pang}, {Paraskeva}, {Perley}, {Petitpas}, {Pursiainen}, {Ravi}, {Ridden-Harper}, {Riddle}, {Rigault}, {Rodriguez}, {Rusholme}, {Sharma}, {Smith}, {Stein}, {Th{\"o}ne}, {Tohuvavohu}, {Valdes}, {van Roestel}, {Vergani}, {Wang}, \& {Zhang}}]{Andreoni2022}
{Andreoni}, I., {Coughlin}, M.~W., {Perley}, D.~A., {et~al.} 2022, \nat, 612, 430, \dodoi{10.1038/s41586-022-05465-8}

\bibitem[{{Astropy Collaboration} {et~al.}(2013){Astropy Collaboration}, {Robitaille}, {Tollerud}, {Greenfield}, {Droettboom}, {Bray}, {Aldcroft}, {Davis}, {Ginsburg}, {Price-Whelan}, {Kerzendorf}, {Conley}, {Crighton}, {Barbary}, {Muna}, {Ferguson}, {Grollier}, {Parikh}, {Nair}, {Unther}, {Deil}, {Woillez}, {Conseil}, {Kramer}, {Turner}, {Singer}, {Fox}, {Weaver}, {Zabalza}, {Edwards}, {Azalee Bostroem}, {Burke}, {Casey}, {Crawford}, {Dencheva}, {Ely}, {Jenness}, {Labrie}, {Lim}, {Pierfederici}, {Pontzen}, {Ptak}, {Refsdal}, {Servillat}, \& {Streicher}}]{2013A&A...558A..33A}
{Astropy Collaboration}, {Robitaille}, T.~P., {Tollerud}, E.~J., {et~al.} 2013, \aap, 558, A33, \dodoi{10.1051/0004-6361/201322068}

\bibitem[{{Astropy Collaboration} {et~al.}(2018){Astropy Collaboration}, {Price-Whelan}, {Sip{\H{o}}cz}, {G{\"u}nther}, {Lim}, {Crawford}, {Conseil}, {Shupe}, {Craig}, {Dencheva}, {Ginsburg}, {VanderPlas}, {Bradley}, {P{\'e}rez-Su{\'a}rez}, {de Val-Borro}, {Aldcroft}, {Cruz}, {Robitaille}, {Tollerud}, {Ardelean}, {Babej}, {Bach}, {Bachetti}, {Bakanov}, {Bamford}, {Barentsen}, {Barmby}, {Baumbach}, {Berry}, {Biscani}, {Boquien}, {Bostroem}, {Bouma}, {Brammer}, {Bray}, {Breytenbach}, {Buddelmeijer}, {Burke}, {Calderone}, {Cano Rodr{\'\i}guez}, {Cara}, {Cardoso}, {Cheedella}, {Copin}, {Corrales}, {Crichton}, {D'Avella}, {Deil}, {Depagne}, {Dietrich}, {Donath}, {Droettboom}, {Earl}, {Erben}, {Fabbro}, {Ferreira}, {Finethy}, {Fox}, {Garrison}, {Gibbons}, {Goldstein}, {Gommers}, {Greco}, {Greenfield}, {Groener}, {Grollier}, {Hagen}, {Hirst}, {Homeier}, {Horton}, {Hosseinzadeh}, {Hu}, {Hunkeler}, {Ivezi{\'c}}, {Jain}, {Jenness}, {Kanarek}, {Kendrew}, {Kern}, {Kerzendorf}, {Khvalko}, {King}, {Kirkby}, {Kulkarni},
  {Kumar}, {Lee}, {Lenz}, {Littlefair}, {Ma}, {Macleod}, {Mastropietro}, {McCully}, {Montagnac}, {Morris}, {Mueller}, {Mumford}, {Muna}, {Murphy}, {Nelson}, {Nguyen}, {Ninan}, {N{\"o}the}, {Ogaz}, {Oh}, {Parejko}, {Parley}, {Pascual}, {Patil}, {Patil}, {Plunkett}, {Prochaska}, {Rastogi}, {Reddy Janga}, {Sabater}, {Sakurikar}, {Seifert}, {Sherbert}, {Sherwood-Taylor}, {Shih}, {Sick}, {Silbiger}, {Singanamalla}, {Singer}, {Sladen}, {Sooley}, {Sornarajah}, {Streicher}, {Teuben}, {Thomas}, {Tremblay}, {Turner}, {Terr{\'o}n}, {van Kerkwijk}, {de la Vega}, {Watkins}, {Weaver}, {Whitmore}, {Woillez}, {Zabalza}, \& {Astropy Contributors}}]{2018AJ....156..123A}
{Astropy Collaboration}, {Price-Whelan}, A.~M., {Sip{\H{o}}cz}, B.~M., {et~al.} 2018, \aj, 156, 123, \dodoi{10.3847/1538-3881/aabc4f}

\bibitem[{{Barniol Duran} {et~al.}(2013){Barniol Duran}, {Nakar}, \& {Piran}}]{BarniolDuran2013}
{Barniol Duran}, R., {Nakar}, E., \& {Piran}, T. 2013, \apj, 772, 78, \dodoi{10.1088/0004-637X/772/1/78}

\bibitem[{{Blandford} \& {K{\"o}nigl}(1979)}]{Blandford1979}
{Blandford}, R.~D., \& {K{\"o}nigl}, A. 1979, \apj, 232, 34, \dodoi{10.1086/157262}

\bibitem[{{Bloom} {et~al.}(2011){Bloom}, {Giannios}, {Metzger}, {Cenko}, {Perley}, {Butler}, {Tanvir}, {Levan}, {O'Brien}, {Strubbe}, {De Colle}, {Ramirez-Ruiz}, {Lee}, {Nayakshin}, {Quataert}, {King}, {Cucchiara}, {Guillochon}, {Bower}, {Fruchter}, {Morgan}, \& {van der Horst}}]{Bloom2011}
{Bloom}, J.~S., {Giannios}, D., {Metzger}, B.~D., {et~al.} 2011, Science, 333, 203, \dodoi{10.1126/science.1207150}

\bibitem[{{Bonnerot} {et~al.}(2016){Bonnerot}, {Rossi}, {Lodato}, \& {Price}}]{Bonnerot2016}
{Bonnerot}, C., {Rossi}, E.~M., {Lodato}, G., \& {Price}, D.~J. 2016, \mnras, 455, 2253, \dodoi{10.1093/mnras/stv2411}

\bibitem[{{Brinkmann} {et~al.}(2000){Brinkmann}, {Laurent-Muehleisen}, {Voges}, {Siebert}, {Becker}, {Brotherton}, {White}, \& {Gregg}}]{Brinkmann2000}
{Brinkmann}, W., {Laurent-Muehleisen}, S.~A., {Voges}, W., {et~al.} 2000, \aap, 356, 445

\bibitem[{{Burrows} {et~al.}(2011){Burrows}, {Kennea}, {Ghisellini}, {Mangano}, {Zhang}, {Page}, {Eracleous}, {Romano}, {Sakamoto}, {Falcone}, {Osborne}, {Campana}, {Beardmore}, {Breeveld}, {Chester}, {Corbet}, {Covino}, {Cummings}, {D'Avanzo}, {D'Elia}, {Esposito}, {Evans}, {Fugazza}, {Gelbord}, {Hiroi}, {Holland}, {Huang}, {Im}, {Israel}, {Jeon}, {Jeon}, {Jun}, {Kawai}, {Kim}, {Krimm}, {Marshall}, {P. M{\'e}sz{\'a}ros}, {Negoro}, {Omodei}, {Park}, {Perkins}, {Sugizaki}, {Sung}, {Tagliaferri}, {Troja}, {Ueda}, {Urata}, {Usui}, {Antonelli}, {Barthelmy}, {Cusumano}, {Giommi}, {Melandri}, {Perri}, {Racusin}, {Sbarufatti}, {Siegel}, \& {Gehrels}}]{Burrows2011}
{Burrows}, D.~N., {Kennea}, J.~A., {Ghisellini}, G., {et~al.} 2011, \nat, 476, 421, \dodoi{10.1038/nature10374}

\bibitem[{{CASA Team} {et~al.}(2022){CASA Team}, {Bean}, {Bhatnagar}, {Castro}, {Donovan Meyer}, {Emonts}, {Garcia}, {Garwood}, {Golap}, {Gonzalez Villalba}, {Harris}, {Hayashi}, {Hoskins}, {Hsieh}, {Jagannathan}, {Kawasaki}, {Keimpema}, {Kettenis}, {Lopez}, {Marvil}, {Masters}, {McNichols}, {Mehringer}, {Miel}, {Moellenbrock}, {Montesino}, {Nakazato}, {Ott}, {Petry}, {Pokorny}, {Raba}, {Rau}, {Schiebel}, {Schweighart}, {Sekhar}, {Shimada}, {Small}, {Steeb}, {Sugimoto}, {Suoranta}, {Tsutsumi}, {van Bemmel}, {Verkouter}, {Wells}, {Xiong}, {Szomoru}, {Griffith}, {Glendenning}, \& {Kern}}]{CASA2022}
{CASA Team}, {Bean}, B., {Bhatnagar}, S., {et~al.} 2022, \pasp, 134, 114501, \dodoi{10.1088/1538-3873/ac9642}

\bibitem[{{Cendes} {et~al.}(2021){Cendes}, {Alexander}, {Berger}, {Eftekhari}, {Williams}, \& {Chornock}}]{Cendes2021}
{Cendes}, Y., {Alexander}, K.~D., {Berger}, E., {et~al.} 2021, \apj, 919, 127, \dodoi{10.3847/1538-4357/ac110a}

\bibitem[{{Cendes} {et~al.}(2022){Cendes}, {Berger}, {Alexander}, {Gomez}, {Hajela}, {Chornock}, {Laskar}, {Margutti}, {Metzger}, {Bietenholz}, {Brethauer}, \& {Wieringa}}]{Cendes2022}
{Cendes}, Y., {Berger}, E., {Alexander}, K.~D., {et~al.} 2022, \apj, 938, 28, \dodoi{10.3847/1538-4357/ac88d0}

\bibitem[{{Cendes} {et~al.}(2024){Cendes}, {Berger}, {Alexander}, {Chornock}, {Margutti}, {Metzger}, {Wieringa}, {Bietenholz}, {Hajela}, {Laskar}, {Stroh}, \& {Terreran}}]{Cendes2024}
---. 2024, \apj, 971, 185, \dodoi{10.3847/1538-4357/ad5541}

\bibitem[{{Cordes} \& {Lazio}(2002)}]{Cordes2002}
{Cordes}, J.~M., \& {Lazio}, T.~J.~W. 2002, arXiv e-prints, astro, \dodoi{10.48550/arXiv.astro-ph/0207156}

\bibitem[{{Dai} {et~al.}(2015){Dai}, {McKinney}, \& {Miller}}]{Dai2015}
{Dai}, L., {McKinney}, J.~C., \& {Miller}, M.~C. 2015, \apjl, 812, L39, \dodoi{10.1088/2041-8205/812/2/L39}

\bibitem[{{Dai} {et~al.}(2018){Dai}, {McKinney}, {Roth}, {Ramirez-Ruiz}, \& {Miller}}]{Dai2018}
{Dai}, L., {McKinney}, J.~C., {Roth}, N., {Ramirez-Ruiz}, E., \& {Miller}, M.~C. 2018, \apjl, 859, L20, \dodoi{10.3847/2041-8213/aab429}

\bibitem[{{Donley} {et~al.}(2002){Donley}, {Brandt}, {Eracleous}, \& {Boller}}]{Donley2002}
{Donley}, J.~L., {Brandt}, W.~N., {Eracleous}, M., \& {Boller}, T. 2002, \aj, 124, 1308, \dodoi{10.1086/342280}

\bibitem[{{Duchesne} {et~al.}(2023){Duchesne}, {Thomson}, {Pritchard}, {Lenc}, {Moss}, {McConnell}, {Wieringa}, {Whiting}, {Wang}, {Wang}, {Rose}, {Raja}, {Murphy}, {Leung}, {Huynh}, {Hotan}, {Hodgson}, \& {Heald}}]{RACs2}
{Duchesne}, S.~W., {Thomson}, A.~J.~M., {Pritchard}, J., {et~al.} 2023, \pasa, 40, e034, \dodoi{10.1017/pasa.2023.31}

\bibitem[{{Duchesne} {et~al.}(2024){Duchesne}, {Grundy}, {Heald}, {Lenc}, {Leung}, {McConnell}, {Murphy}, {Pritchard}, {Rose}, {Thomson}, {Wang}, {Wang}, \& {Whiting}}]{RACs3}
{Duchesne}, S.~W., {Grundy}, J.~A., {Heald}, G.~H., {et~al.} 2024, \pasa, 41, e003, \dodoi{10.1017/pasa.2023.60}

\bibitem[{{Falcke} {et~al.}(2004){Falcke}, {K{\"o}rding}, \& {Markoff}}]{Falcke2004}
{Falcke}, H., {K{\"o}rding}, E., \& {Markoff}, S. 2004, \aap, 414, 895, \dodoi{10.1051/0004-6361:20031683}

\bibitem[{{Fender}(2001)}]{Fender2001}
{Fender}, R.~P. 2001, \mnras, 322, 31, \dodoi{10.1046/j.1365-8711.2001.04080.x}

\bibitem[{{Fender} {et~al.}(2004){Fender}, {Belloni}, \& {Gallo}}]{Fender2004}
{Fender}, R.~P., {Belloni}, T.~M., \& {Gallo}, E. 2004, \mnras, 355, 1105, \dodoi{10.1111/j.1365-2966.2004.08384.x}

\bibitem[{{Foreman-Mackey} {et~al.}(2013){Foreman-Mackey}, {Hogg}, {Lang}, \& {Goodman}}]{Mackey2013}
{Foreman-Mackey}, D., {Hogg}, D.~W., {Lang}, D., \& {Goodman}, J. 2013, \pasp, 125, 306, \dodoi{10.1086/670067}

\bibitem[{{Gallo} {et~al.}(2003){Gallo}, {Fender}, \& {Pooley}}]{Gallo2003}
{Gallo}, E., {Fender}, R.~P., \& {Pooley}, G.~G. 2003, \mnras, 344, 60, \dodoi{10.1046/j.1365-8711.2003.06791.x}

\bibitem[{{Gezari} {et~al.}(2020){Gezari}, {Frederick}, {van Velzen}, {Tartaglia}, {Sollerman}, {Goobar}, {Perley}, \& {Kulkarni}}]{Gezari2020}
{Gezari}, S., {Frederick}, S., {van Velzen}, S., {et~al.} 2020, The Astronomer's Telegram, 13717, 1

\bibitem[{{Gezari} {et~al.}(2009){Gezari}, {Heckman}, {Cenko}, {Eracleous}, {Forster}, {Gon{\c{c}}alves}, {Martin}, {Morrissey}, {Neff}, {Seibert}, {Schiminovich}, \& {Wyder}}]{Gezari2009}
{Gezari}, S., {Heckman}, T., {Cenko}, S.~B., {et~al.} 2009, \apj, 698, 1367, \dodoi{10.1088/0004-637X/698/2/1367}

\bibitem[{{Goodwin} {et~al.}(2022){Goodwin}, {van Velzen}, {Miller-Jones}, {Mummery}, {Bietenholz}, {Wederfoort}, {Hammerstein}, {Bonnerot}, {Hoffmann}, \& {Yan}}]{Goodwin2022}
{Goodwin}, A.~J., {van Velzen}, S., {Miller-Jones}, J.~C.~A., {et~al.} 2022, \mnras, 511, 5328, \dodoi{10.1093/mnras/stac333}

\bibitem[{{Goodwin} {et~al.}(2023{\natexlab{a}}){Goodwin}, {Alexander}, {Miller-Jones}, {Bietenholz}, {van Velzen}, {Anderson}, {Berger}, {Cendes}, {Chornock}, {Coppejans}, {Eftekhari}, {Gezari}, {Laskar}, {Ramirez-Ruiz}, \& {Saxton}}]{Goodwin2023}
{Goodwin}, A.~J., {Alexander}, K.~D., {Miller-Jones}, J.~C.~A., {et~al.} 2023{\natexlab{a}}, \mnras, 522, 5084, \dodoi{10.1093/mnras/stad1258}

\bibitem[{{Goodwin} {et~al.}(2023{\natexlab{b}}){Goodwin}, {Miller-Jones}, {van Velzen}, {Bietenholz}, {Greenland}, {Cenko}, {Gezari}, {Horesh}, {Sivakoff}, {Yan}, {Yu}, \& {Zhang}}]{Goodwin2023b}
{Goodwin}, A.~J., {Miller-Jones}, J.~C.~A., {van Velzen}, S., {et~al.} 2023{\natexlab{b}}, \mnras, 518, 847, \dodoi{10.1093/mnras/stac3127}

\bibitem[{{Goodwin} {et~al.}(2024{\natexlab{a}}){Goodwin}, {Anderson}, {Miller-Jones}, {Malyali}, {Grotova}, {Homan}, {Kawka}, {Krumpe}, {Liu}, \& {Rau}}]{Goodwin2024}
{Goodwin}, A.~J., {Anderson}, G.~E., {Miller-Jones}, J.~C.~A., {et~al.} 2024{\natexlab{a}}, \mnras, 528, 7123, \dodoi{10.1093/mnras/stae362}

\bibitem[{{Goodwin} {et~al.}(2024{\natexlab{b}}){Goodwin}, {Mummery}, {Laskar}, {Alexander}, {Anderson}, {Bietenholz}, {Bonnerot}, {Christy}, {Golay}, {Lu}, {Margutti}, {Miller-Jones}, {Ramirez-Ruiz}, {Saxton}, \& {van Velzen}}]{Goodwin2024b}
{Goodwin}, A.~J., {Mummery}, A., {Laskar}, T., {et~al.} 2024{\natexlab{b}}, arXiv e-prints, arXiv:2410.18665, \dodoi{10.48550/arXiv.2410.18665}

\bibitem[{{Granot} \& {Sari}(2002)}]{Granot2002}
{Granot}, J., \& {Sari}, R. 2002, \apj, 568, 820, \dodoi{10.1086/338966}

\bibitem[{{Grotova} {et~al.}(2025){Grotova}, {Rau}, {Salvato}, {Buchner}, {Goodwin}, {Liu}, {Malyali}, {Merloni}, {Tub\'in-Arenas}, {Homan}, {Krumpe}, {Nandra}, {Shirley}, {Anderson}, {Arcodia}, {Bahic}, {Baldini}, {Buckley}, {Ciroi}, {Kawka}, {Masterson}, {Miller-Jones}, \& {Di Mille}}]{Grotova2025}
{Grotova}, I., {Rau}, A., {Salvato}, M., {et~al.} 2025, \aap, accepted

\bibitem[{Grotova {et~al.}(2025)Grotova, Rau, Baldini, Goodwin, Liu, Merloni, Salvato, Anderson, Arcodia, Buchner, Krumpe, Malyali, Masterson, Miller-Jones, Nandra, \& Shirley}]{Grotova_inpress}
Grotova, I., Rau, A., Baldini, P., {et~al.} 2025, arXiv e-prints, arXiv:2504.08424, \dodoi{10.48550/arXiv.2504.08424}

\bibitem[{{Guillochon} {et~al.}(2014){Guillochon}, {Manukian}, \& {Ramirez-Ruiz}}]{Guillochon2014b}
{Guillochon}, J., {Manukian}, H., \& {Ramirez-Ruiz}, E. 2014, \apj, 783, 23, \dodoi{10.1088/0004-637X/783/1/23}

\bibitem[{{Guillochon} \& {Ramirez-Ruiz}(2015)}]{Guillochon2015}
{Guillochon}, J., \& {Ramirez-Ruiz}, E. 2015, \apj, 809, 166, \dodoi{10.1088/0004-637X/809/2/166}

\bibitem[{{Hajela} {et~al.}(2024){Hajela}, {Alexander}, {Margutti}, {Chornock}, {Bietenholz}, {Christy}, {Stroh}, {Terreran}, {Saxton}, {Komossa}, {Bright}, {Ramirez-Ruiz}, {Coppejans}, {Leung}, {Cendes}, {Wiston}, {Laskar}, {Horesh}, {Schroeder}, {Nayana A.}, {Wieringa}, {Velez}, {Berger}, {Blanchard}, {Eftekhari}, {Gomez}, {Nicholl}, {Sears}, \& {Zauderer}}]{Hajela2024}
{Hajela}, A., {Alexander}, K.~D., {Margutti}, R., {et~al.} 2024, arXiv e-prints, arXiv:2407.19019, \dodoi{10.48550/arXiv.2407.19019}

\bibitem[{{Hammerstein} {et~al.}(2022){Hammerstein}, {van Velzen}, {Gezari}, {Cenko}, \& {ZTF TDE Working Group}}]{Hammerstein2022}
{Hammerstein}, E., {van Velzen}, S., {Gezari}, S., {Cenko}, B., \& {ZTF TDE Working Group}. 2022, in American Astronomical Society Meeting Abstracts, Vol. 240, American Astronomical Society Meeting \#240, 432.07

\bibitem[{{Hayasaki} {et~al.}(2016){Hayasaki}, {Stone}, \& {Loeb}}]{Hayasaki2016}
{Hayasaki}, K., {Stone}, N., \& {Loeb}, A. 2016, \mnras, 461, 3760, \dodoi{10.1093/mnras/stw1387}

\bibitem[{{Ho} \& {Ulvestad}(2001)}]{Ho2001}
{Ho}, L.~C., \& {Ulvestad}, J.~S. 2001, \apjs, 133, 77, \dodoi{10.1086/319185}

\bibitem[{{Homan} {et~al.}(2023){Homan}, {Krumpe}, {Markowitz}, {Saha}, {Gokus}, {Partington}, {Lamer}, {Malyali}, {Liu}, {Rau}, {Grotova}, {Cackett}, {Buckley}, {Ciroi}, {Di Mille}, {Gendreau}, {Gromadzki}, {Krishnan}, {Schramm}, \& {Steiner}}]{Homan2023}
{Homan}, D., {Krumpe}, M., {Markowitz}, A., {et~al.} 2023, \aap, 672, A167, \dodoi{10.1051/0004-6361/202245078}

\bibitem[{{Horesh} {et~al.}(2021{\natexlab{a}}){Horesh}, {Cenko}, \& {Arcavi}}]{Horesh2021}
{Horesh}, A., {Cenko}, S.~B., \& {Arcavi}, I. 2021{\natexlab{a}}, Nature Astronomy, 5, 491, \dodoi{10.1038/s41550-021-01300-8}

\bibitem[{{Horesh} {et~al.}(2021{\natexlab{b}}){Horesh}, {Sfaradi}, {Fender}, {Green}, {Williams}, \& {Bright}}]{Horesh2021b}
{Horesh}, A., {Sfaradi}, I., {Fender}, R., {et~al.} 2021{\natexlab{b}}, \apjl, 920, L5, \dodoi{10.3847/2041-8213/ac25fe}

\bibitem[{{Hovatta} {et~al.}(2008){Hovatta}, {Nieppola}, {Tornikoski}, {Valtaoja}, {Aller}, \& {Aller}}]{Hovatta2008}
{Hovatta}, T., {Nieppola}, E., {Tornikoski}, M., {et~al.} 2008, \aap, 485, 51, \dodoi{10.1051/0004-6361:200809806}

\bibitem[{{Hu} {et~al.}(2024){Hu}, {Price}, \& {Mandel}}]{Hu2024}
{Hu}, F.~F., {Price}, D.~J., \& {Mandel}, I. 2024, \apjl, 963, L27, \dodoi{10.3847/2041-8213/ad29ec}

\bibitem[{{Irwin} {et~al.}(2015){Irwin}, {Henriksen}, {Krause}, {Wang}, {Wiegert}, {Murphy}, {Heald}, \& {Perlman}}]{Irwin2015}
{Irwin}, J.~A., {Henriksen}, R.~N., {Krause}, M., {et~al.} 2015, \apj, 809, 172, \dodoi{10.1088/0004-637X/809/2/172}

\bibitem[{{Kajava} {et~al.}(2020){Kajava}, {Giustini}, {Saxton}, \& {Miniutti}}]{Kajava2020}
{Kajava}, J. J.~E., {Giustini}, M., {Saxton}, R.~D., \& {Miniutti}, G. 2020, \aap, 639, A100, \dodoi{10.1051/0004-6361/202038165}

\bibitem[{{Komossa} \& {Bade}(1999)}]{Komossa1999}
{Komossa}, S., \& {Bade}, N. 1999, \aap, 343, 775, \dodoi{10.48550/arXiv.astro-ph/9901141}

\bibitem[{{Krolik} \& {Chen}(1991)}]{Krolik1991}
{Krolik}, J.~H., \& {Chen}, W. 1991, \aj, 102, 1659, \dodoi{10.1086/115985}

\bibitem[{{Lacy} {et~al.}(2020){Lacy}, {Baum}, {Chandler}, {Chatterjee}, {Clarke}, {Deustua}, {English}, {Farnes}, {Gaensler}, {Gugliucci}, {Hallinan}, {Kent}, {Kimball}, {Law}, {Lazio}, {Marvil}, {Mao}, {Medlin}, {Mooley}, {Murphy}, {Myers}, {Osten}, {Richards}, {Rosolowsky}, {Rudnick}, {Schinzel}, {Sivakoff}, {Sjouwerman}, {Taylor}, {White}, {Wrobel}, {Andernach}, {Beasley}, {Berger}, {Bhatnager}, {Birkinshaw}, {Bower}, {Brandt}, {Brown}, {Burke-Spolaor}, {Butler}, {Comerford}, {Demorest}, {Fu}, {Giacintucci}, {Golap}, {G{\"u}th}, {Hales}, {Hiriart}, {Hodge}, {Horesh}, {Ivezi{\'c}}, {Jarvis}, {Kamble}, {Kassim}, {Liu}, {Loinard}, {Lyons}, {Masters}, {Mezcua}, {Moellenbrock}, {Mroczkowski}, {Nyland}, {O'Dea}, {O'Sullivan}, {Peters}, {Radford}, {Rao}, {Robnett}, {Salcido}, {Shen}, {Sobotka}, {Witz}, {Vaccari}, {van Weeren}, {Vargas}, {Williams}, \& {Yoon}}]{VLASS_paper}
{Lacy}, M., {Baum}, S.~A., {Chandler}, C.~J., {et~al.} 2020, \pasp, 132, 035001, \dodoi{10.1088/1538-3873/ab63eb}

\bibitem[{{Levan} {et~al.}(2011){Levan}, {Tanvir}, {Cenko}, {Perley}, {Wiersema}, {Bloom}, {Fruchter}, {de Ugarte Postigo}, {O'Brien}, {Butler}, {van der Horst}, {Leloudas}, {Morgan}, {Misra}, {Bower}, {Farihi}, {Tunnicliffe}, {Modjaz}, {Silverman}, {Hjorth}, {Th{\"o}ne}, {Cucchiara}, {Cer{\'o}n}, {Castro-Tirado}, {Arnold}, {Bremer}, {Brodie}, {Carroll}, {Cooper}, {Curran}, {Cutri}, {Ehle}, {Forbes}, {Fynbo}, {Gorosabel}, {Graham}, {Hoffman}, {Guziy}, {Jakobsson}, {Kamble}, {Kerr}, {Kasliwal}, {Kouveliotou}, {Kocevski}, {Law}, {Nugent}, {Ofek}, {Poznanski}, {Quimby}, {Rol}, {Romanowsky}, {S{\'a}nchez-Ram{\'\i}rez}, {Schulze}, {Singh}, {van Spaandonk}, {Starling}, {Strom}, {Tello}, {Vaduvescu}, {Wheatley}, {Wijers}, {Winters}, \& {Xu}}]{Levan2011}
{Levan}, A.~J., {Tanvir}, N.~R., {Cenko}, S.~B., {et~al.} 2011, Science, 333, 199, \dodoi{10.1126/science.1207143}

\bibitem[{{Lodato} {et~al.}(2009){Lodato}, {King}, \& {Pringle}}]{Lodato2009}
{Lodato}, G., {King}, A.~R., \& {Pringle}, J.~E. 2009, \mnras, 392, 332, \dodoi{10.1111/j.1365-2966.2008.14049.x}

\bibitem[{{Lu} \& {Bonnerot}(2020)}]{Lu2020}
{Lu}, W., \& {Bonnerot}, C. 2020, \mnras, 492, 686, \dodoi{10.1093/mnras/stz3405}

\bibitem[{{Malyali} {et~al.}(2021){Malyali}, {Rau}, {Merloni}, {Nandra}, {Buchner}, {Liu}, {Gezari}, {Sollerman}, {Shappee}, {Trakhtenbrot}, {Arcavi}, {Ricci}, {van Velzen}, {Goobar}, {Frederick}, {Kawka}, {Tartaglia}, {Burke}, {Hiramatsu}, {Schramm}, {van der Boom}, {Anderson}, {Miller-Jones}, {Bellm}, {Drake}, {Duev}, {Fremling}, {Graham}, {Masci}, {Rusholme}, {Soumagnac}, \& {Walters}}]{Malyali2021}
{Malyali}, A., {Rau}, A., {Merloni}, A., {et~al.} 2021, \aap, 647, A9, \dodoi{10.1051/0004-6361/202039681}

\bibitem[{{Malyali} {et~al.}(2023){Malyali}, {Liu}, {Rau}, {Grotova}, {Merloni}, {Goodwin}, {Anderson}, {Miller-Jones}, {Kawka}, {Arcodia}, {Buchner}, {Nandra}, {Homan}, \& {Krumpe}}]{Malyali2023b}
{Malyali}, A., {Liu}, Z., {Rau}, A., {et~al.} 2023, \mnras, 520, 3549, \dodoi{10.1093/mnras/stad022}

\bibitem[{{Matsumoto} \& {Piran}(2023)}]{Matsumoto2023}
{Matsumoto}, T., \& {Piran}, T. 2023, \mnras, 522, 4565, \dodoi{10.1093/mnras/stad1269}

\bibitem[{{McConnell} {et~al.}(2020){McConnell}, {Hale}, {Lenc}, {Banfield}, {Heald}, {Hotan}, {Leung}, {Moss}, {Murphy}, {O'Brien}, {Pritchard}, {Raja}, {Sadler}, {Stewart}, {Thomson}, {Whiting}, {Allison}, {Amy}, {Anderson}, {Ball}, {Bannister}, {Bell}, {Bock}, {Bolton}, {Bunton}, {Chippendale}, {Collier}, {Cooray}, {Cornwell}, {Diamond}, {Edwards}, {Gupta}, {Hayman}, {Heywood}, {Jackson}, {Koribalski}, {Lee-Waddell}, {McClure-Griffiths}, {Ng}, {Norris}, {Phillips}, {Reynolds}, {Roxby}, {Schinckel}, {Shields}, {Tremblay}, {Tzioumis}, {Voronkov}, \& {Westmeier}}]{RACs1}
{McConnell}, D., {Hale}, C.~L., {Lenc}, E., {et~al.} 2020, \pasa, 37, e048, \dodoi{10.1017/pasa.2020.41}

\bibitem[{{Merloni} {et~al.}(2003){Merloni}, {Heinz}, \& {di Matteo}}]{Merloni2003}
{Merloni}, A., {Heinz}, S., \& {di Matteo}, T. 2003, \mnras, 345, 1057, \dodoi{10.1046/j.1365-2966.2003.07017.x}

\bibitem[{{Metzger}(2022)}]{Metzger2022}
{Metzger}, B.~D. 2022, \apjl, 937, L12, \dodoi{10.3847/2041-8213/ac90ba}

\bibitem[{{Metzger} \& {Stone}(2016)}]{Metzger2016}
{Metzger}, B.~D., \& {Stone}, N.~C. 2016, \mnras, 461, 948, \dodoi{10.1093/mnras/stw1394}

\bibitem[{{Mummery} \& {Balbus}(2020)}]{Mummery2020}
{Mummery}, A., \& {Balbus}, S.~A. 2020, \mnras, 492, 5655, \dodoi{10.1093/mnras/staa192}

\bibitem[{{Mummery} {et~al.}(2024){Mummery}, {van Velzen}, {Nathan}, {Ingram}, {Hammerstein}, {Fraser-Taliente}, \& {Balbus}}]{Mummery2024}
{Mummery}, A., {van Velzen}, S., {Nathan}, E., {et~al.} 2024, \mnras, 527, 2452, \dodoi{10.1093/mnras/stad3001}

\bibitem[{{Pasham} {et~al.}(2023){Pasham}, {Lucchini}, {Laskar}, {Gompertz}, {Srivastav}, {Nicholl}, {Smartt}, {Miller-Jones}, {Alexander}, {Fender}, {Smith}, {Fulton}, {Dewangan}, {Gendreau}, {Coughlin}, {Rhodes}, {Horesh}, {van Velzen}, {Sfaradi}, {Guolo}, {Castro Segura}, {Aamer}, {Anderson}, {Arcavi}, {Brennan}, {Chambers}, {Charalampopoulos}, {Chen}, {Clocchiatti}, {de Boer}, {Dennefeld}, {Ferrara}, {Galbany}, {Gao}, {Gillanders}, {Goodwin}, {Gromadzki}, {Huber}, {Jonker}, {Joshi}, {Kara}, {Killestein}, {Kosec}, {Kocevski}, {Leloudas}, {Lin}, {Margutti}, {Mattila}, {Moore}, {M{\"u}ller-Bravo}, {Ngeow}, {Oates}, {Onori}, {Pan}, {Perez-Torres}, {Rani}, {Remillard}, {Ridley}, {Schulze}, {Sheng}, {Shingles}, {Smith}, {Steiner}, {Wainscoat}, {Wevers}, \& {Yang}}]{Pasham2023}
{Pasham}, D.~R., {Lucchini}, M., {Laskar}, T., {et~al.} 2023, Nature Astronomy, 7, 88, \dodoi{10.1038/s41550-022-01820-x}

\bibitem[{{Piro} \& {Mockler}(2024)}]{Piro2024}
{Piro}, A.~L., \& {Mockler}, B. 2024, arXiv e-prints, arXiv:2412.01922, \dodoi{10.48550/arXiv.2412.01922}

\bibitem[{{Rees}(1988)}]{Rees1988}
{Rees}, M.~J. 1988, \nat, 333, 523, \dodoi{10.1038/333523a0}

\bibitem[{{Roth} {et~al.}(2016){Roth}, {Kasen}, {Guillochon}, \& {Ramirez-Ruiz}}]{Roth2016}
{Roth}, N., {Kasen}, D., {Guillochon}, J., \& {Ramirez-Ruiz}, E. 2016, \apj, 827, 3, \dodoi{10.3847/0004-637X/827/1/3}

\bibitem[{{Sfaradi} {et~al.}(2022){Sfaradi}, {Horesh}, {Fender}, {Green}, {Williams}, {Bright}, \& {Schulze}}]{Sfaradi2022}
{Sfaradi}, I., {Horesh}, A., {Fender}, R., {et~al.} 2022, \apj, 933, 176, \dodoi{10.3847/1538-4357/ac74bc}

\bibitem[{{Shiokawa} {et~al.}(2015){Shiokawa}, {Krolik}, {Cheng}, {Piran}, \& {Noble}}]{Shiokawa2015}
{Shiokawa}, H., {Krolik}, J.~H., {Cheng}, R.~M., {Piran}, T., \& {Noble}, S.~C. 2015, \apj, 804, 85, \dodoi{10.1088/0004-637X/804/2/85}

\bibitem[{{Truemper}(1982)}]{Truemper1982}
{Truemper}, J. 1982, Advances in Space Research, 2, 241, \dodoi{10.1016/0273-1177(82)90070-9}

\bibitem[{{van Velzen} {et~al.}(2020){van Velzen}, {Holoien}, {Onori}, {Hung}, \& {Arcavi}}]{vanVelzen2020}
{van Velzen}, S., {Holoien}, T. W.~S., {Onori}, F., {Hung}, T., \& {Arcavi}, I. 2020, \ssr, 216, 124, \dodoi{10.1007/s11214-020-00753-z}

\bibitem[{{van Velzen} {et~al.}(2016){van Velzen}, {Anderson}, {Stone}, {Fraser}, {Wevers}, {Metzger}, {Jonker}, {van der Horst}, {Staley}, {Mendez}, {Miller-Jones}, {Hodgkin}, {Campbell}, \& {Fender}}]{vanvelzen2016}
{van Velzen}, S., {Anderson}, G.~E., {Stone}, N.~C., {et~al.} 2016, Science, 351, 62, \dodoi{10.1126/science.aad1182}

\bibitem[{{Walker}(1998)}]{Walker1998}
{Walker}, M.~A. 1998, \mnras, 294, 307, \dodoi{10.1046/j.1365-8711.1998.01238.x10.1111/j.1365-8711.1998.01238.x}

\bibitem[{{Wang} {et~al.}(2023){Wang}, {Baldi}, {del Palacio}, {Guolo}, {Yang}, {Zhang}, {Done}, {Castro Segura}, {Pasham}, {Middleton}, {Altamirano}, {Gandhi}, {Qiao}, {Jiang}, {Yan}, {Giroletti}, {Migliori}, {McHardy}, {Panessa}, {Jin}, {Shen}, \& {Dai}}]{Wang2023}
{Wang}, Y., {Baldi}, R.~D., {del Palacio}, S., {et~al.} 2023, \mnras, 520, 2417, \dodoi{10.1093/mnras/stad101}

\bibitem[{{Yao} {et~al.}(2023){Yao}, {Ravi}, {Gezari}, {van Velzen}, {Lu}, {Schulze}, {Somalwar}, {Kulkarni}, {Hammerstein}, {Nicholl}, {Graham}, {Perley}, {Cenko}, {Stein}, {Ricarte}, {Chadayammuri}, {Quataert}, {Bellm}, {Bloom}, {Dekany}, {Drake}, {Groom}, {Mahabal}, {Prince}, {Riddle}, {Rusholme}, {Sharma}, {Sollerman}, \& {Yan}}]{Yao2023}
{Yao}, Y., {Ravi}, V., {Gezari}, S., {et~al.} 2023, \apjl, 955, L6, \dodoi{10.3847/2041-8213/acf216}

\bibitem[{{Yuan} {et~al.}(2015){Yuan}, {Zhang}, {Feng}, {Zhang}, {Ling}, {Zhao}, {Deng}, {Qiu}, {Osborne}, {O'Brien}, {Willingale}, {Lapington}, {Fraser}, \& {the Einstein Probe team}}]{Yuan2015}
{Yuan}, W., {Zhang}, C., {Feng}, H., {et~al.} 2015, arXiv e-prints, arXiv:1506.07735, \dodoi{10.48550/arXiv.1506.07735}

\bibitem[{{Zauderer} {et~al.}(2011){Zauderer}, {Berger}, {Soderberg}, {Loeb}, {Narayan}, {Frail}, {Petitpas}, {Brunthaler}, {Chornock}, {Carpenter}, {Pooley}, {Mooley}, {Kulkarni}, {Margutti}, {Fox}, {Nakar}, {Patel}, {Volgenau}, {Culverhouse}, {Bietenholz}, {Rupen}, {Max-Moerbeck}, {Readhead}, {Richards}, {Shepherd}, {Storm}, \& {Hull}}]{Zauderer2011}
{Zauderer}, B.~A., {Berger}, E., {Soderberg}, A.~M., {et~al.} 2011, \nat, 476, 425, \dodoi{10.1038/nature10366}

\end{thebibliography}
\bibliographystyle{aasjournal}

\end{document}